# Electro-Optic Active Metasurfaces for High-Speed Photonic Applications

*Mingwei Tang[1,2,#,*], Ruochong Chen[1,#], Yue Cao[1], Chao Meng[3], Fei Ding[3,4], Yadong Deng[5], Yubing Han[1], Kai Wei[1,2,*], and Sergey I. Bozhevolnyi[3,*]*

[1] State Key Laboratory of Extreme Photonics and Instrumentation, Zhejiang Key Laboratory of Autonomous Optoelectronic Perception, College of Optical Science and Engineering, Zhejiang University, Hangzhou 310027, China.

[2] ZJU-Hangzhou Global Scientific and Technological Innovation Center, Zhejiang University, Hangzhou 311215, China.

[3] Centre for Nano Optics, University of Southern Denmark, Campusvej 55, Odense DK-5230, Denmark.

[4] School of Electronic Science and Technology, Eastern Institute of Technology, Ningbo 315200, China

[5] Hangzhou Institute of Advanced Studies, Zhejiang Normal University, Hangzhou 311231, China.

* Address all correspondence to:

Mingwei Tang (tangmw@zju.edu.cn), Kai Wei (kwei@zju.edu.cn), Sergey I. Bozhevolnyi (seib@mci.sdu.dk)

**Abstract**

Metasurfaces are artificially engineered ultrathin nanostructured surfaces, capable of flexibly manipulating light–matter interactions on compact platforms, and thereby of great significance for a wide range of applications within modern optics and photonics, including communications, computing, sensing, and quantum technologies. However, the inherently static nature of conventional metasurfaces severely limits their functionalities and thus range of possible applications. Benefiting from integration of the metasurface platform for shaping optical wavefronts with ultrafast electro-optic (EO) materials, active EO metasurfaces have emerged as a frontier research direction targeting advanced photonic devices. This paper systematically reviews the latest progress in this field, featuring a comprehensive comparison of performances and application scenarios of mainstream EO materials such as lithium niobate, barium titanate and organic EO polymers. Modulation mechanisms based on the Pockels and Kerr effects along with the corresponding active metasurface implementations are summarized. Furthermore, improvements in modulation efficiency enabled by advantageously exploiting resonant structural designs and associated phenomena, including Fabry–Pérot resonances, Mie resonances, surface plasmon polaritons, quasi–bound states in the continuum, surface lattice resonances, and guided-mode resonances, are presented and summerized in detail. Current challenges related to metasurface design, nanofabrication, performance and heterogeneous integration are also discussed. Finally, future research directions are outlined, highlighting interdisciplinary developments, novel material engineering, and AI-assisted design as key pathways to enable practical use of active EO metasurfaces in modern optics and photonics, including quantum information technologies.

## 1. Introduction

Over past decades, metasurfaces based on surface micro- and nanostructures have demonstrated remarkable aptitudes for sophisticated optical field manipulations, emerging as powerful platforms for multidimensional control of the amplitude, phase, polarization, and wavefront of electromagnetic waves[1–3]. Contrasting with conventional bulk optical components, metasurfaces rely on ultrathin subwavelength-scale structural units to achieve high-precision optical control and multiple functionalities, significantly reducing device sizes and system complexity. This provides a novel pathway towards miniaturization and integration of high-performance planar optical devices[4–6]. However, most early developed metasurface components were inherently static, meaning that their optical responses were fixed once fabricated and could hardly be reconfigured, which severely limited their applications in adaptive imaging, dynamic displays, and programmable photonic chips.

To overcome this limitation and enable real-time control and modification of optical responses, a wide variety of strategies for realizing dynamic metasurfaces have been proposed[7–12]. According to the nature of external stimuli or material properties, dynamic tuning approaches can generally be classified into two major categories based on (1)mechanical deformation that can be achieved with microelectromechanical systems (MEMS)[13–17]or flexible materials[18,19], or (2)changing refractive index by external-field-driven (electric[20], magnetic[21], optical[22], thermal[23], and chemical[24,25]) modulation in functional materials, such as liquid crystals[21,26–29], phase change materials[30,31], electro-optic (EO) materials[32], two-dimensional (2D) semiconductors[33,34], and doped semiconductors[35,36]. Among these approaches, dynamic metasurfaces based on electrically driven tuning have attracted particular attention due to their aptitude for reaching ultrafast responses and high integration densities while being naturally compatible with electronic control circuits, remarkable features that make electrically controlled metasurfaces very promising for realizing dynamic metasurface functionalities[37].

Amid various electrically tunable materials, EO materials are especially attractive, because their intrinsically ultrafast EO effects enable direct and fast refractive index modulation by applied electric fields, resulting thereby in high-speed, low-loss, continuous, and reversible optical modulation. Since the EO effects originate from asymmetric distortion of outer electron distributions under an external electric field, this electronic response, being practically massless and without structural changes, results in almost instantaneous refractive index changes with applied fields, reaching response speeds exceeding 100 GHz[38]. At the same time, however, one is faced with a formidable challenge when implementing EO metasurfaces: EO induced refractive index changes are intrinsically very small while available interaction lengths in metasurfaces are fundamentally very short, making the task of realizing strong EO modulation of optical fields with metasurfaces extremely difficult. To circumvent this arduous design challenge, one should exploit resonant configurations that would increase the effective interaction length and thereby the modulation efficiency (relative modulation per applied Volt) obtained with even relatively weak EO effects[39]. Once this design problem is satisfactorily dealt with, EO materials become ideal candidates for constructing ultra-fast and efficient dynamic metasurfaces[40].

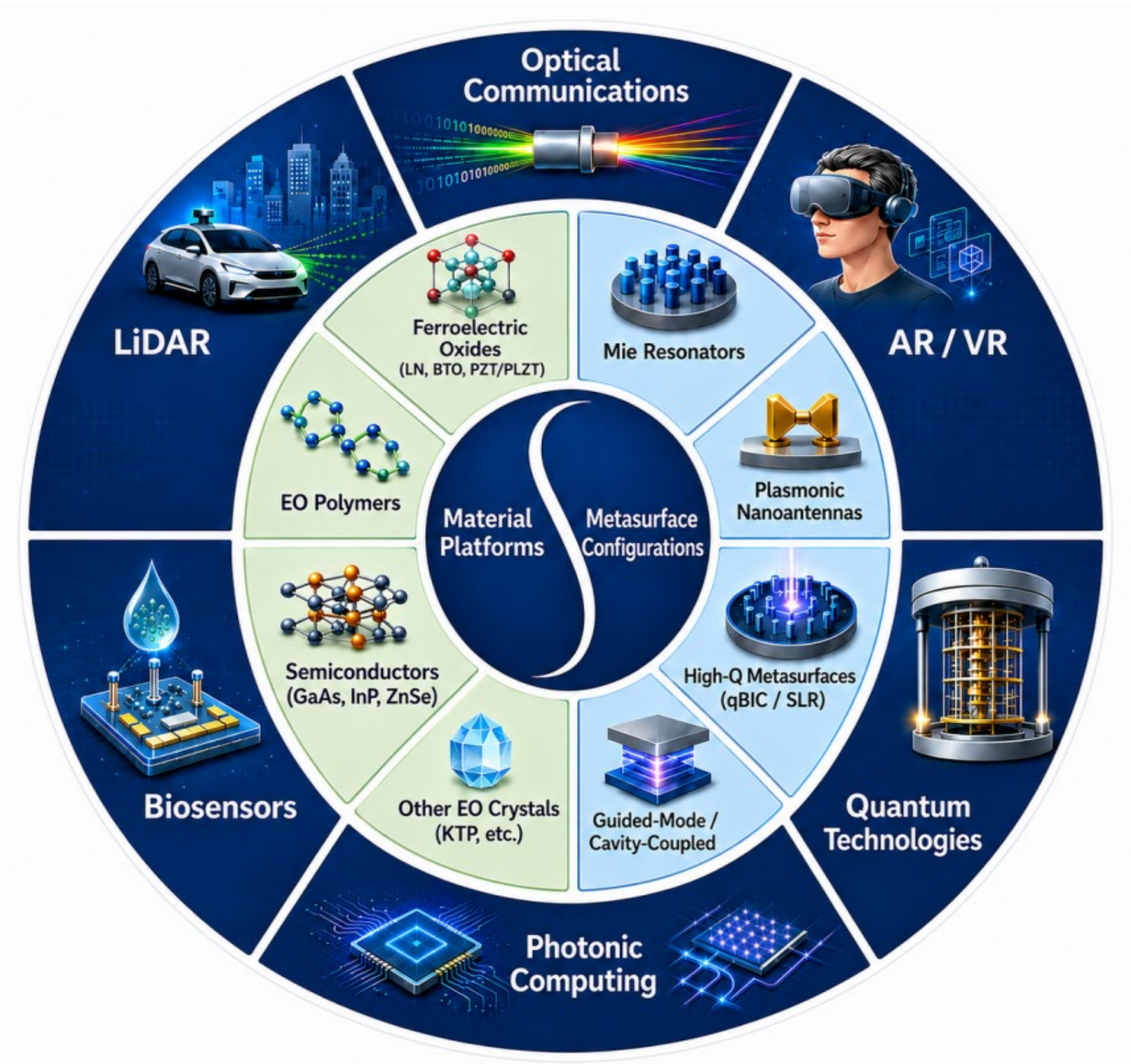

Fig. 1. Schematic illustration of various electro-optic (EO) materials, metasurface configurations and their applications.

Representative EO materials include lithium niobate ($LiNbO_3$, LN), which exhibits a strong Pockels response and a wide transparency window; barium titanate ($BaTiO_3$, BTO), featuring a high dielectric constant and large EO coefficients; organic electro-optic (OEO) polymers with mechanical flexibility and high EO tunability; as well as other inorganic EO crystals such as Potassium titanyl phosphate (KTP)[41] , lead zirconate titanate (PZT)[42] , Lead Lanthanum Zirconate Titanate (PLZT)[43] , and Gallium arsenide (GaAs)[44]. These material platforms provide the fundamental physical basis for transforming metasurfaces from static into high-speed tunable microdevices and systems. It should be noted that EO materials have already been widely used in photonic integrated circuits involving well-confined guided radiation (and thereby ensuring both compactness and long interaction lengths), but rarely in optical systems controlling free-space propagation of optical fields.

The development of compact and reconfigurable free-space optical systems is very important to a wide range of applications, including light detection and ranging (LiDAR)[45–47], free-space optical (FSO) communications[48,49], augmented/virtual reality (AR/VR)[50–54], optical microscope[55–59], optical computing[60–68], and entangled quantum networks[69]. The intensive search for EO metasurfaces configurations enabling high modulation efficiency has eventually resulted in several representative design approaches. According to their underlying field-confinement and resonance-engineering mechanisms, these can broadly be classified into five categories:(1) localized resonance-enhanced EO metasurfaces, in which the EO interaction is reinforced by subwavelength resonant modes, such as Mie and plasmonic resonances in high-index dielectric and metallic nanostructures, respectively; (2) radiation-suppressed high-$Q$ EO metasurfaces, where the suppression of radiative leakage gives rise to sharp resonances, represented primarily by quasi–bound states in the continuums (qBICs) and surface lattice resonances (SLRs); (3) Guided-mode resonances (GMRs) metasurfaces that couple free-space incident light into guided modes supported by periodic photonic structures, thereby achieving efficient optical phase modulation[70]; (4) Fabry–Pérot (FP) cavity-based EO metasurfaces, in which FP resonances enhance strength of light–matter interactions by increasing the interaction length and thus improve (spectral or phase) modulation efficiency; and (5) hybrid metasurfaces that integrate multiple resonant mechanisms and/or material platforms within a single architecture to simultaneously exploit their complementary advantages in enabling strong field confinement, high quality factors ($Q$ factors), efficient loss management, and high modulation efficiency[48,71,72].

In this review, we systematically summarize recent advances in dynamic metasurfaces based on EO materials, starting from the material platforms and underlying physical modulation principles. Emphasis is placed on the structural design strategies, performance optimization methods, and application prospects associated with six representative mechanisms outlined above. Our review provides a comprehensive and systematic reference for this emerging research field, facilitating thereby its further developments and applications in high-speed optical communications, optical sensing, and intelligent optical field manipulation.

## 2. EO Material Platform

A clear understanding of EO modulation requires connecting the underlying field-induced optical responses, the material platforms that provide these responses, and the resulting trade-offs in device performance. According to the specific optical property being modulated, EO modulation can generally be classified into electro-absorption modulation and electro-refraction modulation. Electro-absorption modulation mainly alters the absorption characteristics of a material, *i.e.*, the imaginary part of the complex refractive index, enabling thus the modulation of primarily the optical radiation power. In contrast, electro-refraction modulation directly modifies the refractive properties of the material, namely the real part of complex refractive index, thereby enabling direct phase modulation. When combined with interferometric, resonant, or anisotropic optical structures, such refractive-index modulation can be further converted into amplitude modulation[73,74] and, through electrically tunable birefringence, polarization-state control[75]. Owing to its greater versatility and flexibility in optical field manipulation as well as inherently energy-saving nature, electro-refraction modulation is therefore most widely adopted in practical photonic, including EO device applications.

### 2.1 EO Effects

#### Linear EO Effect (Pockels Effect)

The Pockels effect is one of the key physical mechanisms that enable high-speed EO modulation, with demonstrated bandwidths ranging from tens of gigahertz to beyond 100 GHz in optimized integrated devices[76]. It arises when an external electric field is applied to a non-centrosymmetric EO material, resulting in a linear change in its refractive index. This electric-field-induced refractive index variation provides the physical basis for ultrafast, low-loss, and continuous optical modulation.

When an external electric field $E$ is applied, the refractive index tensor of the crystal is modified, with the resulting change in refractive index $\Delta n$ linearly proportional to the applied electric field. This relationship can be expressed as:

$$\Delta n = -\frac{1}{2} n^3 r_{eff} E \quad (1)$$

where $r_{eff}$ denotes the effective EO coefficient and $E$ is the applied electric field. Physically, this effect originates from the field-induced reorientation of electric dipoles inside the material, which modifies the dielectric tensor of the crystal. Typical EO materials that exhibit a pronounced Pockels effect are LN, BTO, and OEO polymers.

Within the framework of classical electromagnetic theory, the optical response of a material system can often be described by expressing polarization *P* as a tensorial power-series expansion in the components of the strength *E* of an applied optical field:

$$P_i = \boldsymbol{\varepsilon_0}(\chi^{(1)}E+\chi^{(2)}E^2 + \chi^{(3)}E^3 \ + \cdots) \tag{2}$$

Here, $P_i$ denotes the *i*-th Cartesian component ($i$ = *x*, *y*, *z*) of the polarization vector, and $\chi^{(n)}$represents the $n^{th}$ - order electric susceptibility. For centrosymmetric materials, the second-order susceptibility satisfies $\chi^{(2)} = 0$, indicating that second-order nonlinear optical effects, including the Pockels effect, are forbidden. Consequently, the Pockels effect is allowed only in non-centrosymmetric media, because a nonzero linear EO tensor requires broken inversion symmetry.

The variation of the refractive-index ellipsoid coefficient tensor under an applied electric field can be expressed as:

$$\Delta\left(\frac{1}{n^2}\right)_{ij} = \sum_k r_{ijk}\, E_k \tag{3}$$

Eq. (1) represents an approximate expression of Eq. (3) along a specific crystallographic direction. Owing to the intrinsic symmetry of the EO tensor, the third-rank tensor $r_{ijk}$ can be simplified and expressed in a reduced contracted form as $r_{ijk} = r_{hk}$. For materials belonging to the trigonal crystal point group 3m, such as LN, the linear EO coefficient matrix can be expressed as:

$$r_{hk} = \begin{bmatrix} 0 & -r_{22} & r_{13} \\ 0 & r_{22} & r_{13} \\ 0 & 0 & r_{33} \\ 0 & r_{42} & 0 \\ r_{42} & 0 & 0 \\ r_{22} & 0 & 0 \end{bmatrix} \tag{4}$$

Therefore, the above relation can be further simplified and explicitly expanded into the following general form:

$$\begin{gathered}(\frac{1}{n_x^2} - r_{22}E_y + r_{13}E_z)x^2 + (\frac{1}{n_y^2} - r_{22}E_y + r_{13}E_z)y^2 + \\ \left(\frac{1}{n_z^2} - r_{33}E_z\right)z^2 + 2r_{42}E_y yz + 2r_{42}E_x xz + 2r_{22}E_x xy = 1\end{gathered} \tag{5}$$

When the direction of the applied electric field is aligned with the optical axis of the crystal, the above expression can be further simplified as:

$$\left(\frac{1}{n_x^2}+r_{13}E_z\right)x^2+\left(\frac{1}{n_y^2}+r_{13}E_z\right)y^2+\left(\frac{1}{n_z^2}-r_{33}E_z\right)z^2=1 \tag{6}$$

Accordingly, under the condition of a small refractive-index perturbation, the refractive indices along the $x$- and $y$-axes can be approximated as:

$$n_x = n_y \approx n_o - \frac{1}{2}n_o^3 r_{13} E_z \tag{7}$$

where $n_o$ denotes the refractive index of the ordinary wave. Accordingly, the refractive index along the $z$-axis satisfies:

$$n_z \approx n_e - \frac{1}{2}n_e^3 r_{33} E_z \tag{8}$$

where $n_e$ denotes the refractive index of the extraordinary wave. A dynamically varying external electric field applied to the crystal can induce piezoelectric deformation, whose influence on the refractive index mainly depends on the frequency of the electric field. At low frequencies (~kHz), the crystal undergoes strain without stress, and the unclamped EO tensor $r_{hk}^T$ should be employed. In contrast, at high frequencies (~GHz), the inertia of the crystal lattice suppresses lattice deformation, giving rise to stress without strain, and the clamped EO tensor $\mathrm{r}_{hk}^{\mathrm{S}}$ should be used. Since most practical EO modulation applications operate in the high-speed regime, the EO tensor $r_{hk}$ used throughout this work refers to the clamped EO tensor $\mathrm{r}_{hk}^{\mathrm{S}}$.

## 2.2 EO Materials

### Representative EO materials

Representative EO materials exhibit distinct crystal or molecular structures, which strongly influence their EO response, dielectric behavior, optical transparency, and integration compatibility. **Fig. 2** schematically compares the characteristic structural motifs of several representative EO material platforms, including the crystal unit cells of LN, BTO, and PZT, as well as the repeat-unit and dipole-alignment schematic of OEO polymers.

Because the modulation efficiency, optical confinement, electrode design, and thermal stability of EO metasurfaces are strongly influenced by the choice of material, **Table 1** benchmarks representative EO material platforms, specifically LN, BTO, PZT, and OEO polymers, based on their EO coefficients, refractive indices, dielectric permittivities, and characteristic transition temperatures.

**LN** is the most widely used EO material, especially in optical communication systems. It exhibits excellent thermal, chemical, and mechanical stability, and provides a well-balanced combination of electro-optic, optical, and fabrication-related properties. Compared with some EO thin-film materials such as polycrystalline films or EO polymers—whose properties can be strongly affected by crystallinity, domain structure, poling efficiency, strain, or interface quality, thin-film LN can retain many bulk-crystal advantages after nanophotonic processing, including a stable Pockels response, broad transparency, and low optical loss.

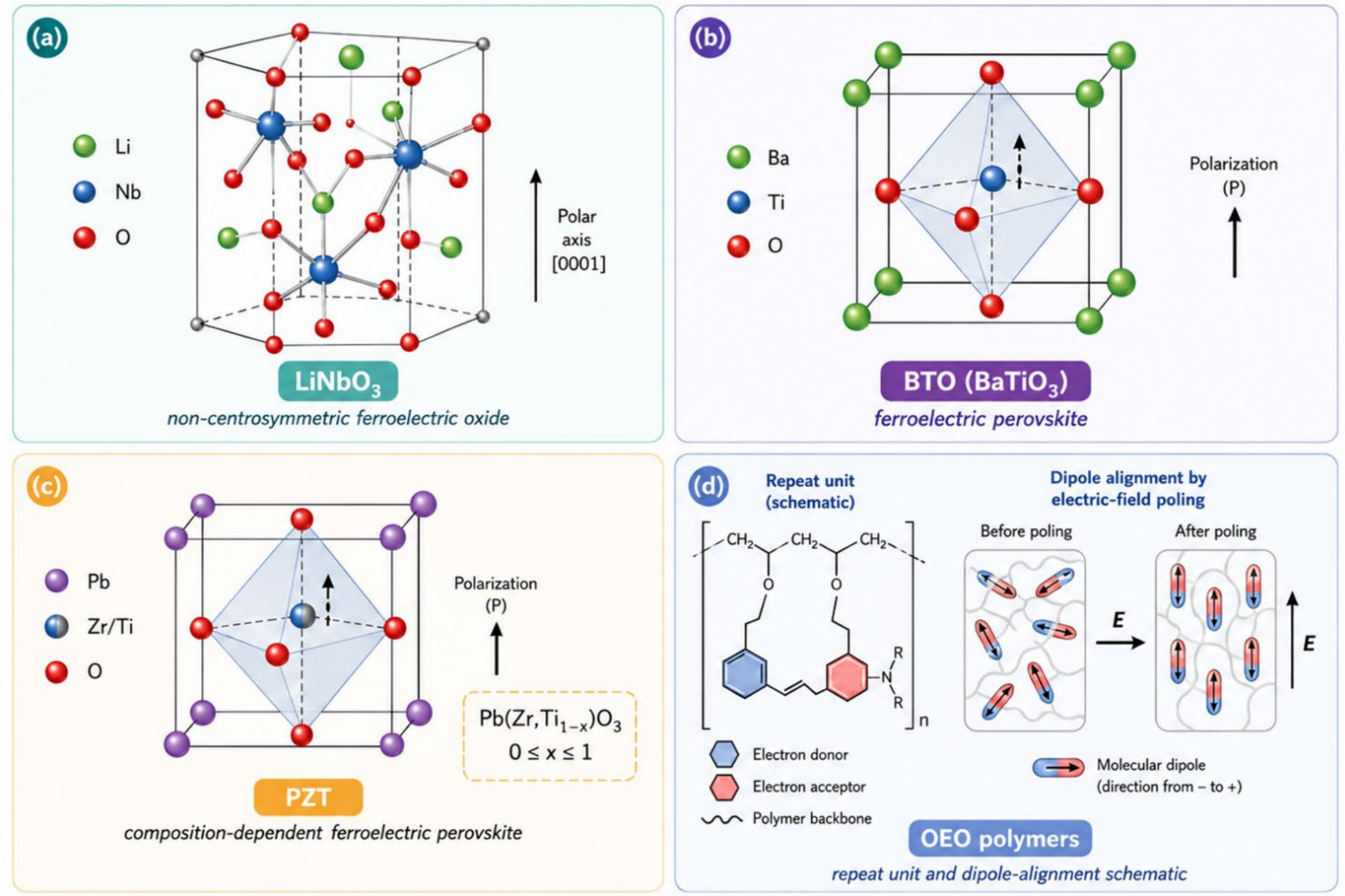

Fig. 2. Representative structural motifs of EO material platforms, *i.e.* LN(a), BTO(b), PZT(c), OEO polymers(d).

Below the Curie temperature, LN[77] exists in a ferroelectric phase with a trigonal crystal structure and a non-centrosymmetric 3m point-group symmetry, as shown in **Fig. 2a**. At a telecom wavelength of $\lambda$=1.55 μm, it exhibits negative birefringence with $n_o$≈2.21, $n_e$≈2.14. It also possesses a wide optical transparency window ranging from 0.35 to 4.5 μm, enabling operation from the visible to the mid-infrared region.

A major drawback of LN is its susceptibility to photorefractive damage. Under illumination at wavelengths around 500 nm with an optical intensity of approximately 20 W/cm², noticeable photodamage will occur[78], which limits its high-power applications in the visible regime[79]. However, for wavelengths exceeding 800 nm, the photorefractive effect becomes negligible[80]; consequently, LN is predominantly employed in the telecom band.

**Table 1**. EO-relevant properties of representative EO material platforms

| EO Materials | EO coefficient (pm/V) | Refractive index | Optical transparency window(μm) | Relative permittivity (unclamped) | Phase-transition temperature $T_g$(°C) | Ref |
|---|---|---|---|---|---|---|
| **LN** | $r_{33}$=31 | $n_o$ = 2.22<br>$n_e$ = 2.14 | 0.35 - 4.5 | $\varepsilon_{33} = 28$<br>$\varepsilon_{11} = 85$ | 1100-1210 | [77,81–85] |
| **BTO** | $r_{42}$=150-1300 | $n_o$ = 2.38<br>$n_e$ = 2.31 | 0.43 – 6.3 | $\varepsilon_{33} = 135$<br>$\varepsilon_{11} = 3600$ | 120~200 | [76,86–94,94] |
| **PZT** | 60–225 | $n_o$ = 2.41<br>$n_e$ = 2.39 | 0.6 - 2.5 | 500 – 12000 (Varying with the ratio of Zr to Ti) | ~140 | [42,43,95–98] |
| **OEO polymers** | 100-1100 | 1.54-2.05 | 0.4-1.6 | $\varepsilon \approx$ 2–7 | 86-103 | [32,99,100,100–105] |

In early LN waveguide modulators, the refractive-index contrast between the core and cladding was relatively small (≈0.05), and the electrodes were positioned far from the optical waveguide. As a result, the half-wave voltage–length product $V_\pi L$ was as high as ~10 V·cm, and the device length typically reached the centimeter scale, leading to large device capacitance and a bandwidth limitation of about 35 GHz. The emergence of thin-film LN has largely addressed these limitations by enabling stronger optical confinement and closer electrode–waveguide integration. By employing dry etching to pattern thin LN films into ridge waveguides, the distance between the electrodes and the waveguide can be significantly reduced to micron-scale (*e.g.*, 3.5 μm)[106], the propagation loss caused by surface roughness can be lowered to ~0.03-0.4 dB/cm[107,108], and the refractive-index contrast can be increased to ~0.7[109]. These improvements enable strong optical confinement and substantially enhance the EO efficiency, yielding a voltage–length product in the range of 1.8–3.0 V·cm[106,110].

To further improve the EO modulation efficiency, or equivalently to reduce the voltage-length product $V_\pi L$, metallic nanostructures supporting surface plasmon polaritons (SPPs) at metal-dielectric interfaces can be introduced. Such hybrid structures enable simultaneous propagation of plasmonic modes and electrical control signals, leading to extremely strong confinement of both optical and electric fields and, consequently, greatly improved modulation efficiency. Among various plasmonic architectures, modulators based on $LiNbO_3$ have achieved exceptional performance, including a directional coupler with $V_{\pi}L$ of 0.3 V·cm[111] and a slow-light Mach-Zehnder modulator reaching a record-low $V_{\pi}L$ of 0.21 V·cm[38] .

**BTO** exhibits an exceptionally high EO activity—approximately one order of magnitude larger than that of LN ($r_{33}$ > 900 pm/V) and its Pockels coefficient remains high even at the nanoscale, making it an attractive candidate for nanoscale and plasmonic EO modulation. At room temperature, BTO has a tetragonal crystal structure with 4mm point-group symmetry (**Fig. 2b**). At its Curie temperature (~120 °C), it undergoes a phase transition from the ferroelectric tetragonal phase to the paraelectric cubic phase, which is centrosymmetric and thus loses its Pockels activity. Similar to LN, BTO exhibits negative birefringence, with $n_o$≈2.41 and $n_e$≈2.36 at λ=0.633 μm[93].

BTO offers several advantages, including high EO activity (with Pockels coefficients significantly exceeding those of LN), good mechanical strength, chemical stability, a high optical damage threshold, and a broad transparency window spanning approximately 0.4–5 μm, as well as good process compatibility with established semiconductor fabrication techniques. It can be epitaxially grown on silicon, and is compatible with complementary metal-oxide-semiconductor (CMOS) manufacturing processes. BTO integrated into photonic and plasmonic structures has demonstrated remarkable performance metrics. For instance, a low $V_{\pi}L$ of 0.23 V·cm and modulation bandwidth exceeding 40 GHz have been reported[112,113], along with data transmission rates surpassing 250 Gb/s[114].

However, the EO activity of currently available BTO thin films is significantly lower than that of bulk crystals and exhibits strong dependence on film composition and growth conditions. Therefore, further optimization of the growth process, crystallinity, and stoichiometry is required to preserve the excellent properties of bulk BTO for large-scale integrated applications.

**PZT** thin-film is another promising candidate for EO platforms[95] ( **Fig. 2c**). It exhibits good optical transparency across the wavelength range of approximately 0.6 - 2.5 μm[97] and potentially high EO activity, with Pockels coefficients ranging from 60 pm/V for polycrystalline thin films to 225 pm/V for bulk or epitaxial materials, depending on the deposition process and crystalline quality. However, maintaining high Pockels coefficients in PZT thin films is challenging[96]; large values require high-quality perovskite-structured films typically grown at elevated temperatures (typically >600 °C)[97], although lower-temperature routes (down to ~450 °C) have been demonstrated[115]. While metallic intermediate layers are commonly employed as bottom electrodes to facilitate perovskite phase formation, modern processing techniques also enable CMOS-compatible PZT integration[97]. Recently, nanophotonic modulators based on PZT thin films were demonstrated on silicon nitride (SiN) substrates, achieving voltage–length products as low as 1.2 V·cm[42].

**OEO polymers** are typically doped with organic molecules possessing permanent dipole moments and strong nonlinear optical responses. Compared with inorganic EO materials, OEO polymers with extended π-electron conjugation afford the potential for the greatest bandwidth (**Fig. 2d**). The phase-relaxation time of the π-electron system is tens of femtoseconds, which can translate to potential bandwidths of tens of terahertz. During fabrication, a host polymer containing nonlinear optical chromophores is usually dissolved in a solvent and spin-coated onto a substrate, forming a polymer film with randomly oriented chromophores. To induce efficient second-order nonlinearity, an electric poling process is performed at a temperature close to the polymer's glass-transition temperature $T_g$, followed by rapid cooling to well below $T_g$. This process permanently aligns the non-centrosymmetric chromophores along the applied electric field, thereby enabling the Pockels effect.

Efficient electric poling remains a critical challenge because OEO polymers generally possess relatively low dielectric strength. Excessive bias voltage can easily induce large leakage currents and cause dielectric breakdown, leading to irreversible damage to the polymer film and device structure. Another major limitation is the poor thermal stability of OEO polymers. The glass-transition temperatures of practical EO polymers typically lie in the range of 50-150 °C, and the chromophore alignment can gradually relax at temperatures even 30 °C below $T_g$, resulting in a continuous degradation or even complete loss of second-order nonlinearity[92].

Nevertheless, once successfully poled, OEO polymers can exhibit Pockels coefficients as high as 100-400 pm/V, far exceeding that of LN. Modulators based on the silicon-organic hybrid platform have demonstrated voltage-length products well below 1 V·cm[116,117]. In highly efficient plasmonic modulators based on OEO polymers, metal-insulator-metal (MIM) slot waveguides are used to modulate the supported SPPs[118,119], enabling extremely strong light-matter interaction. Record-high modulation performance has been achieved with voltage-length products below 0.1 V·cm and 3 dB bandwidths exceeding 500 GHz[120]. However, the major drawback of such plasmonic devices is their high insertion loss (up to 8 dB), which often requires resonant switching schemes, such as ring-resonator-based modulators, to mitigate Ohmic losses[121].

**Other EO Materials**

**KTP** exhibits a relatively small EO coefficient ($r_{33}$≈36 pm/V, comparable to that of LN), and therefore lower EO activity. In addition, KTP crystal growth requires stringent environmental conditions. However, KTP offers good chemical stability, high mechanical strength, and excellent thermal conductivity, making it suitable for high-power laser modulation.

**La-doped PZT (PLZT)** is formed by doping PZT with La, and its EO performance can be tailored by adjusting the material composition. PLZT exhibits relatively high EO activity (EO coefficient: 200 to 300 pm/V), good optical transparency over a wide spectral range, and stable chemical properties. However, the presence of lead causes environmental concerns, limiting its large-scale industrial use. In addition, the high dielectric permittivity of PLZT increases the device capacitance and radio frequency (RF) loading, which can limit the modulation bandwidth and complicate high-frequency electrode design for telecommunication applications.

**III–V compound semiconductors** such as GaAs exhibit modest EO coefficients, with the only non-zero Pockels component $r_{41}$≈1.5 pm/V at telecom wavelengths[122]. The direct bandgap of 1.42 eV imposes a short-wavelength absorption edge at ~870 nm, precluding visible-band operation; however, GaAs offers an exceptionally broad transparency window from ~0.9 to ~16 μm. Although $r_{41}$ is substantially smaller than the $r_{33}$ of LN, the high refractive index of GaAs (~3.37 at 1.55 μm) partially compensates through the $n^3r$ scaling of the modulator figure of merit. More importantly, the near-coincidence of the optical group index (~3.55) and the RF refractive index ($\sqrt{\varepsilon} \approx 3.6$) facilitates natural velocity matching in traveling-wave electrode designs—a key advantage for broadband operation that has enabled 50 GHz GaAs Mach–Zehnder modulators with $V_\pi L \approx 8.3$ V·cm[123]. The direct bandgap also confers radiation hardness and environmental stability, making GaAs a natural candidate for space-borne systems. Compatibility with mature 6-inch III–V foundry processes, together with favorable thermal and electrical conductivities, further renders it attractive for integrated optoelectronic platforms.

## 2.3 Comparison between EO Materials

EO materials typically exhibit large Pockels coefficients, and the response time of the Pockels effect generally reaches the femtosecond to nanosecond scale[124]. Therefore, the core advantage of EO materials lies in their ability to achieve ultrafast optical phase modulation. In addition, some EO materials also possess outstanding properties such as high Curie temperatures, good chemical stability, and excellent mechanical robustness, which further enhance their practical applicability.

It should be noted, however, although the Pockels effect enables rapid tuning of the refractive index, the refractive-index tuning range remains extremely small. Consequently, resonant structures are commonly introduced into modulators to enhance the light–matter interaction. Typical approaches include exploiting multiple reflections within FP resonances at the metasurface layer[125], or the nonlocal interaction between incident light and guided modes propagating along the metasurface to increase the effective interaction length and accumulate the phase shift induced by the small refractive-index variation[74]. Such resonance-enhanced schemes, however, are usually achieved at the expense of a reduced operational wavelength bandwidth.

**Table 2**. Performance comparison of different EO modulators

| Material platform | Theoretical bandwidth (GHz) | Propagation loss (dB/cm) | Modulation efficiency ($V_{\pi}L$) (V·cm) | Types of modulators | Ref. |
|---|---|---|---|---|---|
| **LN** | 100 | 0.4 | 2.2 | Mach–Zehnder interferometer (MZI) | [110] |
| **LN** | 9(800GHz) | 3500 | 0.21 | Plasmonic modulator | [126] |
| **BTO** | 65/30 | 14000/10 | 0.45 | Plasmonic modulator/ Ring resonator | [92] |
| **BTO** | 12 | / | 0.7 | MZI | [86] |
| **PZT** | 33 | 1 | 3.2 | Ring resonator | [95] |
| **PZT** | >40 | 1.8 | 0.8 | Photonic crystal | [98] |
| **OEO polymers** | 70 | 4000 | 0.006 | Plasmonic modulator | [120] |
| **OEO polymers** | 76 | 10 | 0.13 | MZI | [127] |

To facilitate a systematic comparison among various EO materials, we summarize the performance of optical modulators—the simplest device architecture employing EO materials[73,128] provides a visual comparison of major EO material platforms, illustrating the trade-offs among modulation speed, optical loss, modulation efficiency, Curie temperature, and optical transparency window. Detailed performance parameters for each platform are tabulated in **Table 2**.

Building on the material platform comparison in **Section 2.2** along with the performance metrics summarized above, LN emerges as a platform that achieves an excellent balance among modulation efficiency, propagation loss, and long-term stability. OEO polymers combine exceptionally large Pockels coefficients (up to ~1100 pm/V) with outstanding modulation efficiency; however, their relatively high optical loss (~1 dB/cm) and limited thermal stability remain critical challenges[99]. Titanate perovskites, exemplified by BTO, represent a highly promising material class characterized by high efficiency, moderate loss, and broad bandwidth—although further optimization of thin-film growth processes is essential to preserve the superior EO properties of their bulk counterparts[86].

At present, LN remains the dominant material in commercial optical communication systems and is widely regarded as one of the most promising platforms for next-generation high-performance nanophotonic devices. The rapid advancement of quantum photonics, artificial neural networks, and high-speed communication technologies is driving the evolution of future nanophotonic EO modulators toward device architectures and material platforms that enable highly efficient, ultrafast operation within compact footprints. Consequently, novel materials, such as BTO, PZT, and OEO polymer with larger Pockels coefficients—capable of overcoming the intrinsic EO activity limitations of LN—are of significant interest.

## 3. Operating Mechanisms and Structural Designs of EO Metasurfaces

Compared with conventional bulky EO devices, metasurfaces utilize artificial subwavelength-scale structures to precisely tailor the local electromagnetic field distribution and optical modes, thereby enabling pronounced modulation of phase, amplitude, and polarization within an ultrathin form factor. However, due to the intrinsically limited refractive-index modulation of EO materials, it is often difficult to achieve sufficiently large modulation depth and high device efficiency by relying solely on the intrinsic EO effect. Although the EO effect features an ultrafast response and linear tunability, the achievable refractive-index modulation in practical devices is typically on the order of $10^{-4}$-$10^{-3}$, which severely restricts the modulation efficiency in phase control, beam steering, and intensity modulation.

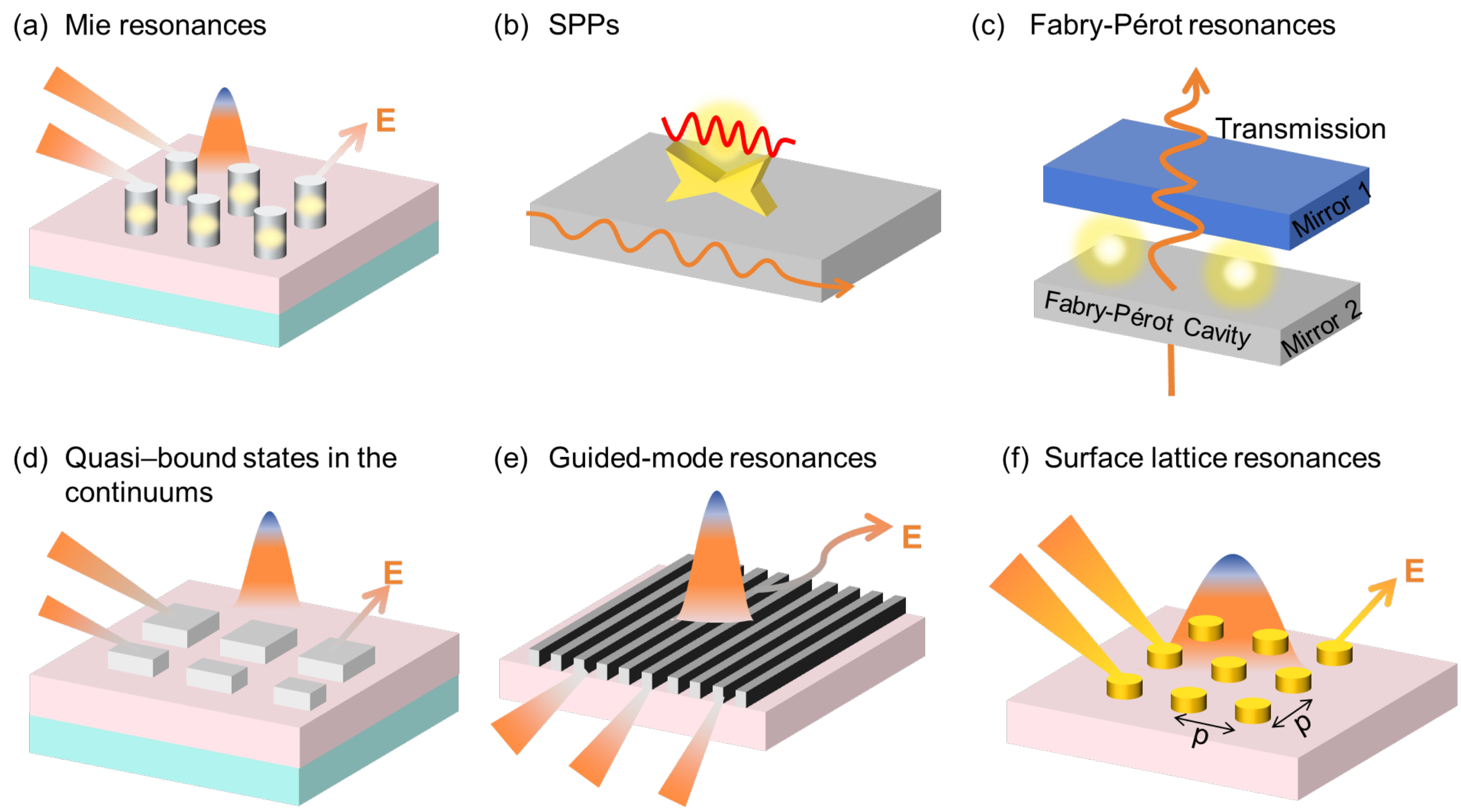

Fig. 3. Schematic illustration of various optical mode-engineering for EO metasurfaces, such as Mie resonances (a), SPPs (b), FP resonances (c), qBICs (d), GMRs (e), SLRs (f).

As illustrated in **Fig. 3**, to overcome the intrinsically limited refractive-index tunability of EO materials, a variety of optical mode-engineering strategies have been developed to enhance light-matter interactions in EO metasurfaces. The core concept is to construct specific resonant modes within subwavelength structures to enhance the residence time of the optical field inside the EO material, thereby amplifying the influence of the EO effect on light. From a physical-mechanism perspective, existing EO metasurfaces can be systematically classified according to the dominant optical resonance and interaction mechanisms employed to amplify the EO response.

A major category of EO metasurfaces relies on localized resonance enhancement, wherein optical modes—such as Mie resonances in high-index dielectric nanoresonators or localized surface plasmon resonances in metallic nanoantennas—are leveraged to confine electromagnetic fields at the nanoscale, thereby enhancing EO modulation at specific wavelengths. However, the quality factors ($Q$ factors) of such localized resonances are typically moderate due to radiative losses or absorption loss, which constrain achievable modulation efficiency and spectral selectivity.

To overcome this limitations, radiation-suppressed high-$Q$ EO metasurfaces have been developed by engineering the coupling between resonant modes and free-space radiation channels. Representative implementations include metasurfaces based on qBICs and SLRs. qBICs arise from perturbations of ideal BICs—either by breaking the symmetry that decouples a symmetry-protected BIC from radiation channels, or by slightly detuning the parameter condition that sustains destructive interference among radiation channels—yielding ultranarrow linewidth resonances with strongly enhanced near-field intensities. In contrast, SLRs originate from the coherent far-field coupling of localized resonances in periodic nanoparticle arrays near Rayleigh anomalies (RAs), combining high $Q$ factors with relatively large mode volumes. These attributes render radiation-suppressed metasurfaces particularly effective for enhancing EO modulation efficiency.

A third important class comprises guided-mode-coupled EO metasurfaces, in which metasurface elements are coupled to guided modes supported by thin-film waveguides. Through momentum matching enabled by periodic modulation, incident light efficiently excites GMRs with reduced radiative loss. The resulting nonlocal optical modes enable enhanced light–matter interaction over extended propagation lengths, facilitating large phase modulation with relatively low optical loss—a characteristic particularly attractive for integrated EO modulation and wavefront-control applications.

Cavity-enhanced EO metasurfaces constitute a fourth category, realized by incorporating FP cavities or reflective layers above and below the metasurface. Multiple round-trip propagation within the cavity effectively prolongs the interaction time between the optical field and the EO material, thereby enhancing modulation depth and contrast. Such cavity-assisted architectures provide a versatile strategy for high-performance EO switching, filtering, and intensity modulation.

Finally, hybrid EO metasurfaces integrate plasmonic nanoantennas, dielectric resonators, and optical cavities within a unified platform. By synergistically combining the ultrastrong near-field enhancement of plasmonic resonances with the low-loss, high-$Q$ characteristics of dielectric or cavity modes, these hybrid architectures achieve EO modulation efficiencies that surpass those attainable using any single enhancement mechanism in isolation.

Collectively, these EO metasurface platforms represent distinct yet complementary strategies for enhancing light–matter interactions. In the following sections, we review prototypical device architectures and physical models for each category, with emphasis on their modulation mechanisms, performance metrics, and practical design considerations.

To quantitatively describe resonance-enhanced EO modulation, the enhancement can be understood from both field localization and resonance lifetime (spectral sharpness). For a resonant nanophotonic mode, the local density of states inside the EO medium can be approximated by the Purcell factor[129],

which scales as:

$$F_p \approx \frac{3}{4\pi^2}\left(\frac{\lambda}{n}\right)^3 \frac{Q}{V_{eff}} \tag{9}$$

where $Q$ is the quality factor, $V_{\rm eff}$ is the effective mode volume, $\lambda$ is the free-space wavelength, and $n$ is the refractive index of the EO medium. This relation indicates that a high-$Q$ resonance with a small $V_{eff}$ concentrates optical energy in the active region and strengthens light–matter interaction, thereby amplifying the impact of an electrically induced refractive-index change $\Delta n$[130]. In addition, EO modulation in resonant metasurfaces is commonly implemented by operating on the steep spectral slope of the resonance, where a small resonance shift produces a large intensity change. Under the small-signal approximation, the transmission (or reflectance) modulation can be expressed as:

$$\Delta T \approx \frac{dT}{d\omega}\frac{d\omega}{dn}\Delta n \tag{10}$$

implying that $|\Delta T|$ is enhanced by a large spectral slope $\left|\frac{dT}{d\omega}\right|$ and a strong index sensitivity $\left|\frac{d\omega}{dn}\right|$. For sharp resonances, the linewidth $\Delta\omega \sim \frac{\omega_0}{Q}$ becomes narrow, leading to a larger $\left|\frac{dT}{d\omega}\right|$ and thus a higher modulation efficiency that approximately scales with $Q$ factors[131].

Although the modulation efficiency can be improved in a resonance-enhanced EO modulator, the modulation bandwidth is fundamentally constrained by the high $Q$. The overall small-signal modulation bandwidth ($f_{\rm EO}$) is governed by two independent factors: the RC-limited bandwidth ($f_{\rm RC}$) and the photon-lifetime-limited bandwidth ($f_\tau$). The relationship is given by[132]:

$$\frac{1}{f_{EO}^2} = \frac{1}{f_{RC}^2} + \frac{1}{f_\tau^2} \tag{11}$$

The photon-lifetime-limited bandwidth is directly determined by the $Q$ factor of the optical resonance:

$$f_\tau = \frac{1}{2\pi\tau} = \frac{c}{Q\lambda} \tag{12}$$

where $\tau$ is the photon lifetime, $c$ is the speed of light, and $\lambda$ is the resonant wavelength.

Hence, a higher $Q$ factor leads to a longer photon lifetime ($\tau$), which inherently reduces $f_\tau$. Consequently, while a high-$Q$ resonance enhances modulation efficiency (*e.g.*, larger extinction ratio or lower driving voltage), it imposes a fundamental upper limit on the achievable modulation bandwidth. To achieve both high efficiency and fast speed, one must carefully balance $Q$ against the RC time constant, ensuring that $f_\tau$ and $f_{\rm RC}$ are comparable and sufficiently high for the target application.

### 3.1 Localized Resonance-Enhanced EO Metasurfaces

Localized resonance-enhanced EO metasurfaces mainly involve the following representative mechanisms:

(1) Mie resonances, where high-index dielectric nanostructures support electric and magnetic dipole modes, enabling strong electric-field confinement and efficient phase control;

(2) SPPs, which arise from the collective oscillation of free electrons and provide ultra-strong electromagnetic field confinement at the nanoscale, thereby enhancing the effective EO modulation at specific resonant wavelengths. These resonance mechanisms can dramatically enhance light–matter interaction in EO metasurfaces by increasing the local light field intensity within the EO material. Benefiting from their strong field confinement and relatively simple structural implementations, Mie- and plasmonic-resonance-based EO metasurfaces have been extensively investigated and widely adopted in early demonstrations of resonantly enhanced EO modulation.

**Mie Resonances**

By exciting electric dipole, magnetic dipole, and higher-order multipolar modes, metasurfaces based on Mie resonances provide an ideal platform for achieving low-loss and strongly localized optical field manipulation[133]. Compared with metallic plasmonic resonances, Mie resonances offer distinct advantages, including rich modal diversity, negligible Ohmic loss, and high design flexibility, which makes them a key carrier for EO modulation.

In recent years, the development of Mie-resonance-based EO metasurfaces has followed a clear evolutionary path from fundamental dipolar mode modulation toward composite higher-order resonant engineering. Early studies, such as that by Weigand et al.[134], exploited the electric-dipole Mie resonance in LN nanopillars (**Fig. 4**a). By operating at the steep slope of the transmission spectrum and utilizing refractive-index modulation, they achieved an effective EO response under a low driving voltage (<1 $V_{pp}$). Although the modulation depth was limited (0.01% at 10 $V_{pp}$), this work verified the feasibility of Mie-resonance-enhanced EO modulation.

Subsequently, research efforts shifted toward achieving performance breakthroughs through structural innovation. Wang et al.[135] systematically compared multiple configurations, including single-disk, double-disk, and membrane–disk hybrid structures (**Fig. 4b**). They found that the double-disk–based Huygens-like resonance enabled a phase modulation of approximately 70° while maintaining high transmission, significantly surpassing the intrinsic limitations of single-mode Mie resonances and clearly demonstrating the advantages of multimode cooperative design.

To further enhance modulation efficiency, recent work by Babicheva et al.[136] explored the $Sn_2P_2S_6$ material platform, which exhibits an exceptionally large EO coefficient (**Fig. 4c**). By exciting higher-order Mie modes such as magnetic octupoles, the degree of field localization was enhanced by approximately 40%, leading to a pronounced spectral shift of 35 nm and a modulation efficiency about five times higher than that of conventional LN-based systems. This advance not only confirms the unique role of higher-order multipolar modes in strengthening light–matter interaction, but also highlights the tremendous potential of combining material innovation with resonance-mode optimization.

Another work using Mie resonance to enhance electro-optic modulation in polycrystalline BTO metasurfaces through optimizing electric field confinement through embedded and conformal device designs. It achieves up to threefold higher modulation than prior work, enables ferroelectric poling for a 25% further increase, and demonstrates MHz-speed, low-voltage operation, offering a scalable CMOS-compatible platform for active metasurfaces[137].

The phase tunability covering $2\pi$ is required for subwavelength wavefront engineering. Transmissive metasurfaces using high-Q Mie resonances with a single resonance and two ports suffer from a transmission zero that restricts phase tuning to ≤180°. Kim et al. introduces additional diffraction channels in reflection to increase the port count, thereby lifting this constraint and enabling continuous 0–360° phase modulation in transmission with spectrally flat amplitude (theoretical bound 0.5)[138]. A step-by-step design methodology based on temporal coupled-mode theory is presented, employing a LN pedestal to engineer the direct scattering parameter. Full-wave simulations validate the approach using Ge (~250° shift, mid-IR) and Si pillars (~300° shift, telecom), both with near-constant transmittance close to the theoretical limit.

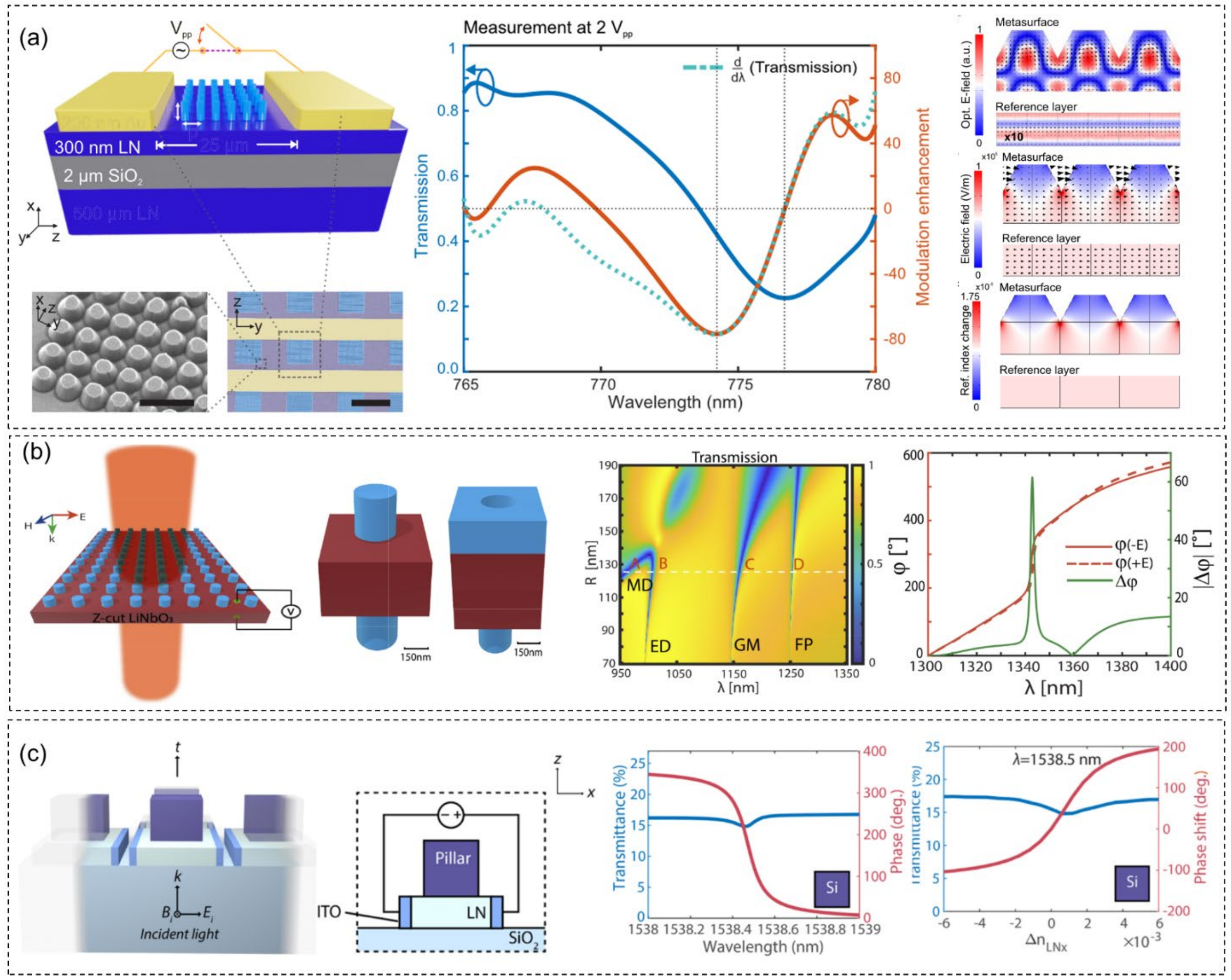

Fig. 4. EO metasurfaces using Mie resonances. (a) Schematic of the LN pillar metasurface and the corresponding distributions of the optical field, electric field, and refractive index within a unit cell. Reproduced with permission[134]. Copyright 2021 American Chemical Society. (b) Schematic diagrams of three representative structures, where the blue regions denote silicon and the red regions denote LN. From left to right: single-disk, double-disk, and membrane-disk structures. Reproduced with permission. The largest transmission phase tunability with relatively high amplitude can be achieved in a dual-disk structure, where the phase can be modulated in a 70 degrees range. Reproduced with permission[135]. Copyright 2022 Optical Society of America. (c) Dynamic near $2\pi$ phase control with a single resonance and a constant transmission. Reproduced with permission[138]. Copyright 2026 American Chemical Society.

Overall, Mie-resonance-based EO metasurfaces have evolved from basic dipolar-mode operation toward performance scaling enabled by composite resonant designs and higher-order multipolar excitations. Both amplitude and phase enhancement could be achieved using Mie-resonance metasurface structures. Looking forward, by coupling Mie resonances with higher-$Q$ modes such as qBICs, it is expected that one can further break the current limits of modulation depth and efficiency while preserving the intrinsic advantages of low loss and high design flexibility, thereby providing a solid technological route toward next-generation high-performance integrated photonic devices.

**SPPs**

SPPs arise from the collective oscillation of free electrons in metallic nanostructures and are characterized by extreme subwavelength field confinement and strongly enhanced local electromagnetic fields[139,140]. Owing to their ability to dramatically amplify light–matter interaction, SPPs provide a powerful physical mechanism for boosting EO modulation efficiency within ultracompact footprints. Compared with dielectric resonances, plasmonic modes intrinsically suffer from Ohmic loss, yet their unparalleled field localization and broadband response make them particularly attractive for high-speed and nanoscale EO devices.

Early demonstrations of SPPs-based EO modulation in LN focused on traveling-wave plasmonic architectures. Thomaschewski et al.[141] employed plasmonic phase shifters integrated with LN to realize an ultracompact Mach–Zehnder modulator based on the Pockels effect (**Fig. 5a**). By using gold nanostripes on z-cut LN to simultaneously guide SPPs and the driving electrical signal, strong subwavelength confinement and a large overlap between optical and electrostatic fields were achieved. This hybrid plasmonic-EO design enabled an exceptionally low half-wave voltage–length product of $V_{\pi}L \approx 0.21\ V \cdot cm$, representing the highest modulation efficiency reported for LN MZMs. Although plasmonic propagation loss and dielectric breakdown across nanoscale gaps limited the achievable extinction ratio (~2.5 dB), this work convincingly demonstrated the ability of SPPs to drastically shrink device footprint while enhancing EO efficiency. The similar plasmonic phase shifters were also arranged to form plasmonic optical phased array (OPA) that can perform aliasing-free beam steering with an angular range of ±5° and flat frequency response up to 18 GHz[142].

Subsequent efforts extended SPPs-based modulation concepts toward resonant plasmonic architectures to further improve efficiency and stability. Recent work[143] demonstrated a plasmonic LN micro-racetrack EO modulator operating on the steep slope of a high-$Q$ resonant cavity (**Fig. 5b**). By combining the strong Pockels nonlinearity of OEO with nanoscale plasmonic confinement, efficient phase-to-intensity conversion was realized within a compact resonator footprint. The device exhibited EO bandwidths exceeding 176 GHz together with enhanced thermal robustness, outperforming conventional resonant Si modulators in operating-point stability. Despite the additional plasmonic absorption loss, this work clearly shows that resonant SPPs enhancement can effectively relax the trade-off between footprint, speed, and modulation efficiency.

More recently, SPPs-empowered EO modulation has been further advanced through material–platform innovation. Kohli et al.[144] developed a plasmonic BTO-on-SiN platform to achieve ultrahigh-speed EO modulation (**Fig. 5c**). By exploiting the exceptionally large Pockels coefficient of BTO in combination with nanoscale plasmonic slot waveguides, extreme optical and electrical field confinement was achieved, enabling efficient phase modulation within device lengths of only a few micrometers. This platform supported electrical bandwidths exceeding 100 GHz and enabled record symbol rates beyond 200 GigaBaud (GBd), including 256 GBd Mach–Zehnder, 224 GBd IQ, and 200 GBd racetrack modulators. Although plasmonic losses remain non-negligible, this work establishes plasmonic–ferroelectric integration as a viable route toward compact, ultrafast EO modulation compatible with large-scale SiN photonic circuits.

Besides traveling-wave plasmonic architectures, localized surface plasmon resonance (LSPR)[145] has also been leveraged to enhance the EO effect. Weiss et al.[83] combined aluminum nanodisk arrays with thin-film LN to form a hybrid structure supporting the coupling of SPPs, SLR, and FP cavity modes (**Fig. 5d**). Under a driving voltage of 50 $V_{pp}$, the device exhibited a resonance tuning range of 3 nm and a modulation depth of 40%, demonstrating the feasibility of FP-cavity-assisted EO modulation based on multi-resonance coupling.

Overall, SPPs-based EO modulators have evolved from early traveling-wave plasmonic phase shifters toward resonant architectures and advanced ferroelectric–plasmonic hybrid platforms. Through synergistic optimization of plasmonic confinement, resonant enhancement, and material nonlinearity, SPPs have demonstrated their unique capability to push EO modulation toward unprecedented speed and compactness. Looking forward, further integration of SPPs-based designs with low-loss photonic routing, improved thermal management, and hybrid resonant concepts may provide a promising pathway toward next-generation high-performance EO modulators that balance efficiency, bandwidth, and scalability.

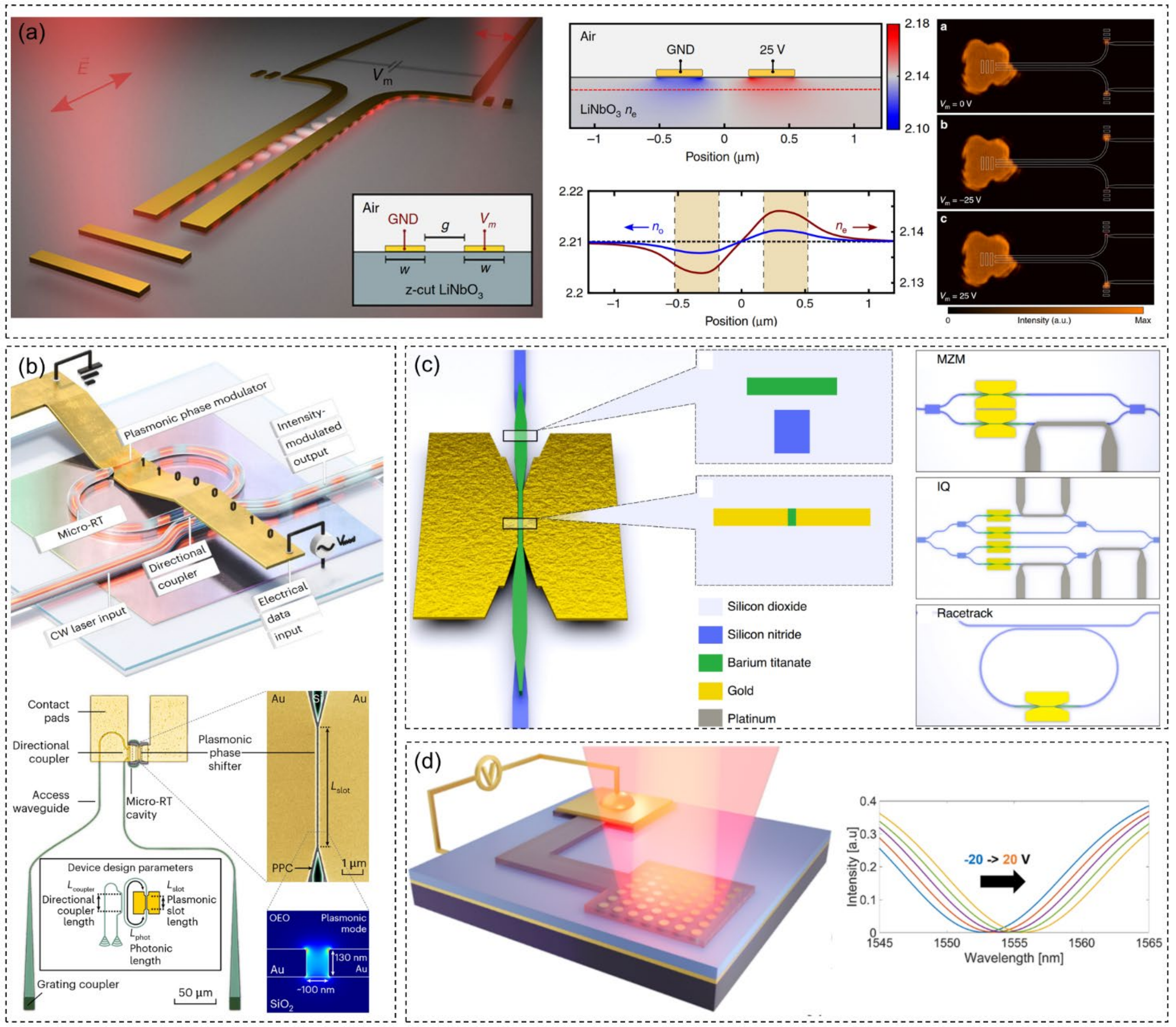

Fig. 5. EO metasurfaces using surface plasmon polaritons (SPPs). (a) Schematic of a plasmonic LN unbalanced Mach–Zehnder modulator, together with the corresponding device geometry and distributions of the optical field, applied electrical field, and Pockels-effect-driven refractive-index modulation, illustrating the strong spatial overlap between the confined plasmonic mode and the electro-optically active region. Reproduced with permission[141]. Copyright 2020 Nature Publishing Group. (b) Schematic and transmission spectra of a plasmonic micro-racetrack modulator, where plasmonic–organic-hybrid phase modulation inside a resonant feedback loop enables efficient phase-to-intensity conversion with broadband, thermally stable, high-speed intensity modulation and large optical modulation amplitude. In the resonant feedback loop, the light is phase modulated within the gold (Au) plasmonic slot waveguide. Reproduced with permission[143]. Copyright 2023 Nature Publishing Group. (c) Conceptual overview of a plasmonic BTO-on-SiN platform, highlighting the integration of low-loss SiN photonics, the strong Pockels nonlinearity of barium titanate, and nanoscale plasmonic confinement to realize ultracompact, high-speed Mach–Zehnder, IQ, and racetrack EO modulators. Reproduced with permission[144]. Copyright 2025 Nature Publishing Group. (d) EO metasurface based on localized surface plasmon resonance (LSPR). Reproduced with permission[83]. Copyright 2022 American Chemical Society.

### 3.2 FP Cavity-Resonant EO Metasurfaces

By using metasurfaces as cavity mirrors or embedding them within the cavity, FP designs can simultaneously offer high field enhancement and additional degrees of freedom for wavefront engineering.

In an FP cavity, multiple reflections between two partially reflecting mirrors lead to constructive interference when the round-trip phase satisfies:

$$2\phi = 2kL = 2\frac{2\pi}{\lambda}nL = 2\pi m \tag{13}$$

where $k$ is the propagation constant, $L$ is the cavity length, $n$ is the effective refractive index, $\lambda$ is the free-space wavelength, and $m$ is an integer mode index[146]. The resulting spectral response follows the Airy function (with individual resonances closely approximating a Lorentzian line shape), and the resonance sharpness is commonly quantified by the free spectral range (FSR) and finesse:

$$\Delta\lambda_{FSR} \approx \frac{\lambda_0^2}{2nL}, \qquad \mathcal{F} = \frac{\Delta\lambda_{FSR}}{\delta\lambda_{FWHM}} \approx \frac{\pi\sqrt{R}}{1-R} \tag{14}$$

where $R$ denotes the effective mirror reflectivity (for a symmetric cavity).

Equivalently, the $Q$ factor can be written as:

$$Q = \frac{\lambda_0}{\delta\lambda_{FWHM}} = \frac{\omega_0}{\Delta\omega} \tag{15}$$

indicating that high-reflectivity mirrors and low internal loss yield narrow linewidths and long photon lifetimes. This combination enables EO modulation schemes in which the resonance line shape and the phase/amplitude response can be jointly tailored.

**Metal film or grating:** The active EO metasurface was first proposed in 2021, an active Fresnel lens using a continuous semitransparent gold film was demonstrated[147], achieving the focusing efficiency of 15% and modulation efficiency of 1.5% (for the driving voltage of ±10 V) within the bandwidth of ~6.4 MHz. This configuration is simple but faces fabrication constraints on film thickness (**Fig. 6a**). Later in 2022, the same group overcomes this limitation by introducing a forked nanostripe electrode configuration[148], which eliminates the need for an ultrathin continuous film. This innovation significantly improves modulation depth of ~20% while maintaining MHz bandwidth, demonstrating a 2×2 individually addressable array and paving the way toward ultrafast spatial light modulators (**Fig. 6b**).

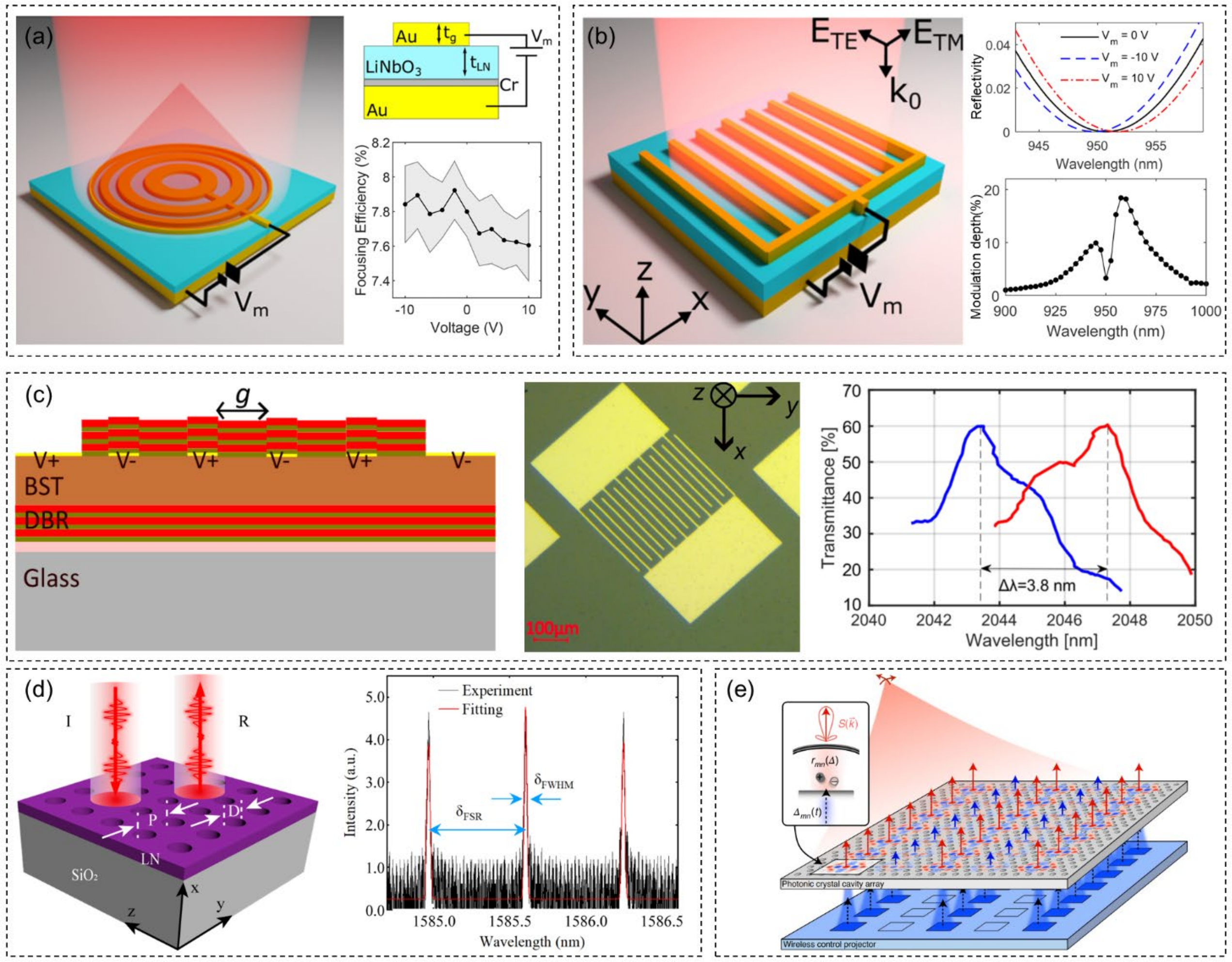

Fig. 6. EO metasurfaces using FP resonance. (a) Schematic of the active Fresnel lens and the focusing modulation of the Fresnel lens. Reproduced with permission[147]. Copyright 2021 American Chemical Society. (b) FP cavity formed by nanostripe metasurface. Reproduced with permission[148]. Copyright 2022 The Royal Society of Chemistry. (c) FP cavity formed by distributed Bragg reflector (DBR) mirrors. Reproduced with permission[149], under a Creative Commons Attribution 4.0 International License.(d) FP cavity formed by photonic crystal (PhC) cavities. Reproduced with permission[150], under a Creative Commons Attribution 4.0 International License. (e) Inverse-designed 2D PhC cavities for full degree-of-freedom spatiotemporal modulation. Reproduced with permission[151]. Copyright 2022 Nature Publishing Group.

**Distributed Bragg reflector (DBR):** To achieve a higher *Q* factor, the DBR mirror can be used to construct the FP cavity. A tunable optical bandpass filter based on a $Ba_{0.5}Sr_{0.5}TiO_3$ thin film integrated into a FP cavity was demonstrated, operating at 2 μm wavelength[149]. The filter achieves 67% peak transmittance and 3.8 nm spectral tuning at 210 V, establishing BST as a viable platform for tunable integrated photonics (**Fig. 6c**).

**Photonic crystal cavities (PhC)**: Recent studies have progressed from establishing high-reflectivity cavity components to demonstrating electrically tunable modulation functions. Liu et al.[150] fabricated a Fano-resonant metasurface on LNOI, achieving a reflectance of 92%. They also integrate the metasurface as a mirror in a FP cavity, achieving a finesse of 38 (**Fig. 6d**). In the future, the FP mode can also be combined with inverse-designed PhC[151] to realize more complex functions, *e.g.* full-degree-of-freedom spatiotemporal light modulator (**Fig. 6e**).

Overall, FP-cavity-assisted EO metasurfaces are moving from basic cavity construction toward performance optimization through hybrid integration and coupled-resonance engineering. The ability to independently design cavity parameters and metasurface responses makes this platform promising for compact modulators with improved modulation contrast and scalable wavefront control. Future work may further benefit from EO materials with larger coefficients and cavity designs that minimize optical loss while maintaining strong field enhancement, enabling higher-speed and lower-voltage operation in reconfigurable photonic systems.

### 3.3 Guided-Mode-Coupled EO Metasurfaces

GMRs enable free-space excitation of leaky waveguide modes in periodic photonic structures, producing spectrally sharp resonances with enhanced optical fields and long photon lifetimes[152]. Up to now, various GMR structures have been demonstrated to achieve ultrahigh $Q$ factors[153,154], the highest up to $2.39 \times 10^5$, which is comparable to the largest $Q$-factor obtained by topological engineering. In general, a GMR is excited when the in-plane wavevector of the incident light, assisted by a reciprocal lattice vector of the grating, satisfies the phase-matching condition[155]:

$$\beta(\lambda) = k_{\parallel} + mG \tag{16}$$

where $\beta = n_{eff} k_0$ is the propagation constant of the guided mode, $k_{\parallel} = k_0 n_{inc} sin\theta$ is the in-plane component of the incident wavevector, $G = 2\pi/\Lambda$ is the grating momentum, $\Lambda$ is the period, m is the diffraction order, and $k_0 = 2\pi/\lambda$[152,156].

When the guided mode couples to the radiation continuum through the periodic perturbation, the resonance linewidth is governed by the total loss rate $\gamma = \gamma_{rad} + \gamma_{abs}$, yielding the quality factor:

$$Q = \frac{\omega_0}{2\gamma} \tag{17}$$

which directly determines the spectral sharpness and photon lifetime. Because GMRs are supported by nonlocal propagating modes rather than purely localized resonances, their spectral response is highly sensitive to refractive-index perturbations, making them attractive for EO modulation and dynamic wavefront control.

Recent progress in GMR-based EO metasurfaces has moved from maximizing the $Q$ factor to balancing field enhancement with modulation speed. In an early study, Klopfer et al.[157] reported GMR with $Q$ factors up to 30,000 in a Si-LN heterogeneous platform, enabling electrically controlled beam steering and beam splitting and highlighting the potential of GMRs for reconfigurable wavefront manipulation(**Fig. 7a**). The simulated result shows that a nearly $2\pi$phase variation can be achieved with an applied bias not exceeding ±25V while keeps the reflection efficiency above 91% at the telecommunication band. Beyond high-$Q$ operation, more recent work has explored engineered modal symmetries to access enhanced field localization. For example, Zhang et al.[158] introduced symmetry breaking in silicon nanorod dimers to excite a magnetic toroidal-dipole GMR, achieving $Q$>5,000 and a magnetic-field enhancement approaching $2.4\times10^4$.

In parallel, device-level demonstrations have increasingly emphasized modulation efficiency and bandwidth. Zheng et al.[103] designed a silicon–organic slot metasurface with optical confinement in a 100-nm-wide slot, yielding a transmission extinction ratio of 38% with a modulation bandwidth of 3 MHz (**Fig. 7b**). A further increase in modulation speed was demonstrated by Dagli et al.[159] using a heterogeneously integrated Si-LN metasurface with GMR. This device maintained a high $Q$ over 1,000 while achieving a transmission modulation depth of 7.1% and a modulation bandwidth exceeding 1 GHz, indicating the feasibility of GMR-based metasurfaces for high-speed EO modulation (**Fig. 7d**). The modulation efficiency can be further increased with an optimized Fano-type qBIC resonance in a LN metasurface platform by breaking the translational symmetry along the direction parallel to the 1D crystal axis (**Fig. 7c**). Francescantonio et al.[85] demonstrated the platform can achieve a modulation efficiency exceeding 10%, driven by less than 10 V and an operational bandwidth of about 1 GHz.

In the forward direction, it's desirable to achieve GHz-rate modulation and megapixel-scale spatial control for commercial applications. Trajtenberg-Mills et al. propose the LN-on-silicon (LNoS) spatial light modulator architecture, integrating a thin-film lithium niobate guided-mode resonance ($Q$ > 1000, field overlap ~90%) with a commercial megapixel CMOS backplane, achieving a 1.6 GHz modulation bandwidth (detector-limited). They employ interference lithography for large-area PhC fabrication and a cold bonding process, demonstrating the first heterogeneous integration of an EO thin film with a CMOS backplane, thereby establishing a scalable pathway toward high-speed, large-aperture SLMs.

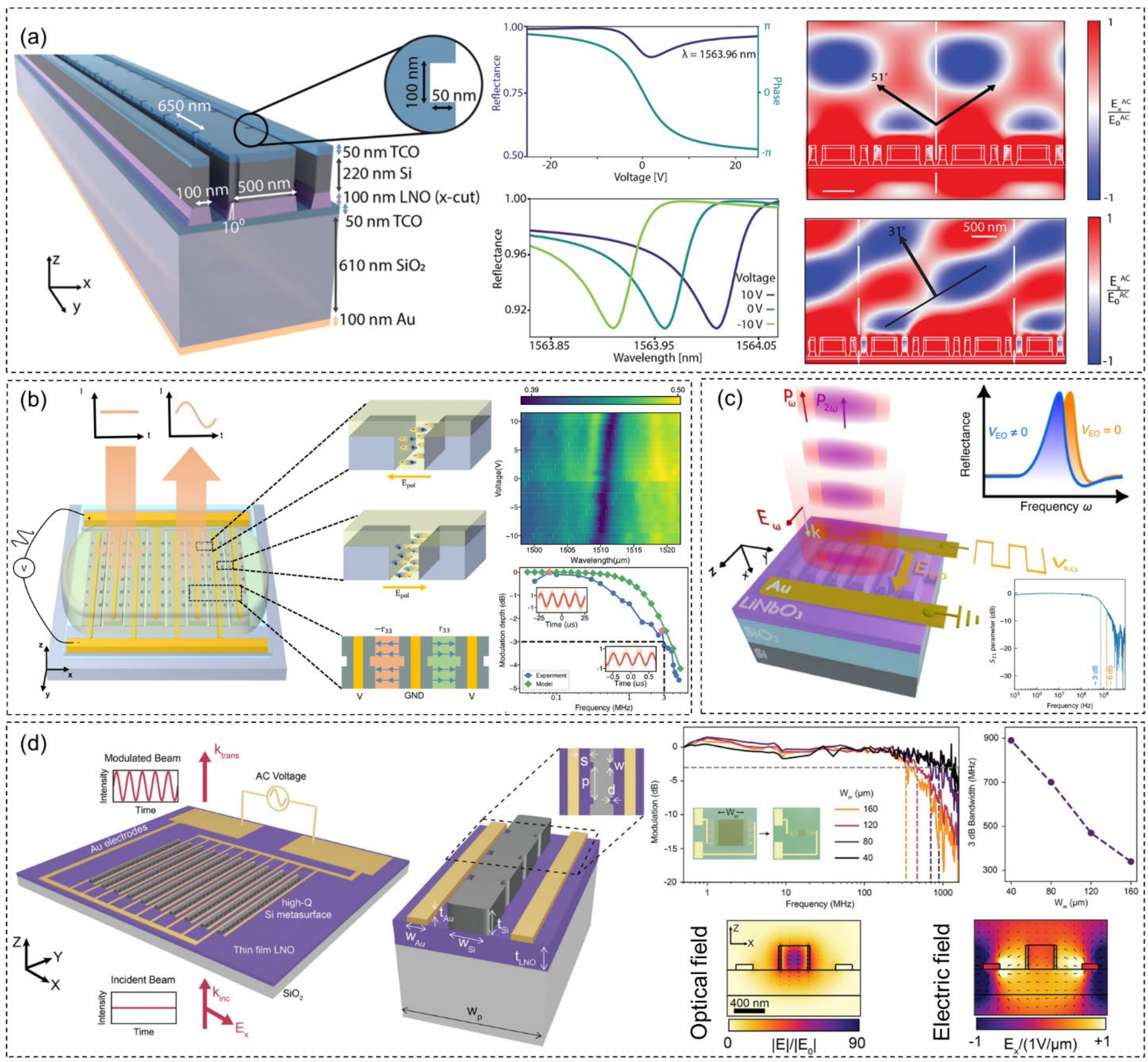

Fig. 7. EO metasurfaces using GMRs. (a) High-Q Si-LN metasurfaces. Reproduced with permission[157]. Copyright 2022 American Chemical Society. (b) Silicon–organic slot metasurface. Reproduced with permission[103]. Copyright 2024 Nature Publishing Group. (c) qBIC LN metasurface for linear and nonlinear optical modulation. Reproduced with permission[85]. Copyright 2025 Nature Publishing Group. (d) GHz-speed wavefront shaping LN metasurface modulators. Reproduced with permission[159]. Copyright 2025 John Wiley and Sons.

Overall, GMR-based EO metasurfaces have transitioned from *Q*-factor-oriented designs to a co-optimization framework involving resonance line shape, field localization, and modulation bandwidth. Their nonlocal mode nature and designable dispersion provide a useful link between localized resonances and extended waveguide photonics[74,160]. Future efforts combining GMRs with higher-field-confinement mechanisms such as qBICs[161] may further improve modulation efficiency while preserving high-speed operation, enabling compact programmable photonic devices.

### 3.4 Radiation-Suppressed High-$Q$ EO Metasurfaces

To further enhance EO modulation efficiency beyond this limit, radiation-suppressed high-$Q$ EO metasurfaces have been developed by engineering the radiation loss, *i.e.*, coupling between resonant modes and free-space radiation channels at the metasurface level.

Unlike localized resonances that primarily rely on field confinement within individual nanostructures, radiation-suppressed high-$Q$ metasurfaces exploit interference and symmetry engineering to suppress radiative leakage, thereby significantly extending the residence time of optical fields[162]. Within this category, two representative mechanisms have been extensively explored. One is based on qBICs, which emerge from perturbations of ideal BICs—either by breaking the symmetry that decouples a symmetry-protected BIC from radiation channels, or by slightly detuning the parameter condition that sustains destructive interference among radiation channels[74]. The other relies on SLRs[163], where coherent far-field coupling in periodic nanostructure arrays near RAs leads to interference-induced suppression of radiation losses and the formation of collective high-$Q$ modes.

By effectively minimizing radiative dissipation, radiation-suppressed high-$Q$ EO metasurfaces provide a powerful route toward highly efficient, spectrally selective EO modulation, and have therefore attracted increasing attention in recent years.

**qBICs**

qBICs are realized by introducing controlled symmetry breaking into originally symmetric structures, thereby transforming ideal BICs with theoretically infinite $Q$ factors into ultrahigh-$Q$ resonances that can be excited from free space. This unique property makes qBICs extremely sensitive to refractive-index perturbations in the surrounding medium, providing an ideal physical platform for low-power and high-efficiency EO modulation. In symmetry-protected BICs and qBIC metasurfaces, the radiative loss can be strongly suppressed, and the quality factor typically follows：

$$Q \propto \frac{1}{\beta^2} \tag{18}$$

with $\beta$ denoting a symmetry-breaking perturbation parameter, providing a practical route to ultrahigh-$Q$ resonances in open metasurface platforms[163,164].

Recent studies have convincingly demonstrated the great potential of this approach. Benea-Chelmus et al. [104]demonstrate a hybrid silicon-organic metasurface platform for GHz-speed free-space electro-optic modulation by combining silicon Mie resonators supporting qBICs (Q up to 550 at λ ~ 1594 nm) with high-performance JRD1 organic molecules ($r_{33}$ = 100 pm/V), as shown in **Fig. 8a**. Key results include DC resonance tuning of $\Delta\lambda$ = 11 nm for qBIC modes (surpassing the linewidth), a modulation depth of $\Delta T/T_{max}$ = 67% at $V_{switch}$ = 100 V, and RF modulation up to 5 GHz ($f_{-3dB}$ = 3 GHz). GMRs achieve even larger tuning of $\Delta\lambda$ = 20 nm. The design decouples optical resonator engineering from electrode geometry, and the post-fabrication spin-coating of the active layer enables versatile integration with arbitrary nanostructure geometries and materials.

Kanyang et al.[165] designed an electrically tunable Huygens' qBIC metasurface based on LN, as shown in **Fig. 8b**. By introducing structural asymmetry to excite electric- and magnetic-dipole qBIC modes, they realized nearly $2\pi$ continuous phase modulation while maintaining high transmission efficiency, enabling excellent wavefront shaping performance. Subsequently, Shen et al.[166]exploited qBIC resonances in a BTO metasurface and achieved 100% modulation depth under a low driving voltage of only 14.3 V, clearly highlighting the strong amplification of tiny refractive-index changes enabled by qBICs. In 2025, Damgaard-Carstensenet al.[74] further integrated a metallic nanograting onto a LN thin film (**Fig. 8c**), where qBIC excitation in the telecommunication band enabled a modulation depth of 95% together with a 125 MHz bandwidth, providing a promising solution for high-speed free-space optical communication.

Compared with conventional Mie resonances, the ultranarrow spectral linewidth and ultrahigh $Q$ factor of qBICs allow for a substantial reduction in the driving voltage and a marked improvement in modulation efficiency, thus offering superior EO control performance[134]. The highest $Q$ of $1.2 \times 10^5$ has been demonstrated in TFLN platform based on photonic crystal microcavities[167]. Looking forward, the combination of the extreme sensitivity of qBICs with the rich modal controllability of Mie resonances, together with the incorporation of anisotropic media and emerging material platforms such as two-dimensional materials[168–171], is expected to open new avenues for the development of next-generation programmable, low-power, and ultrafast photonic devices.

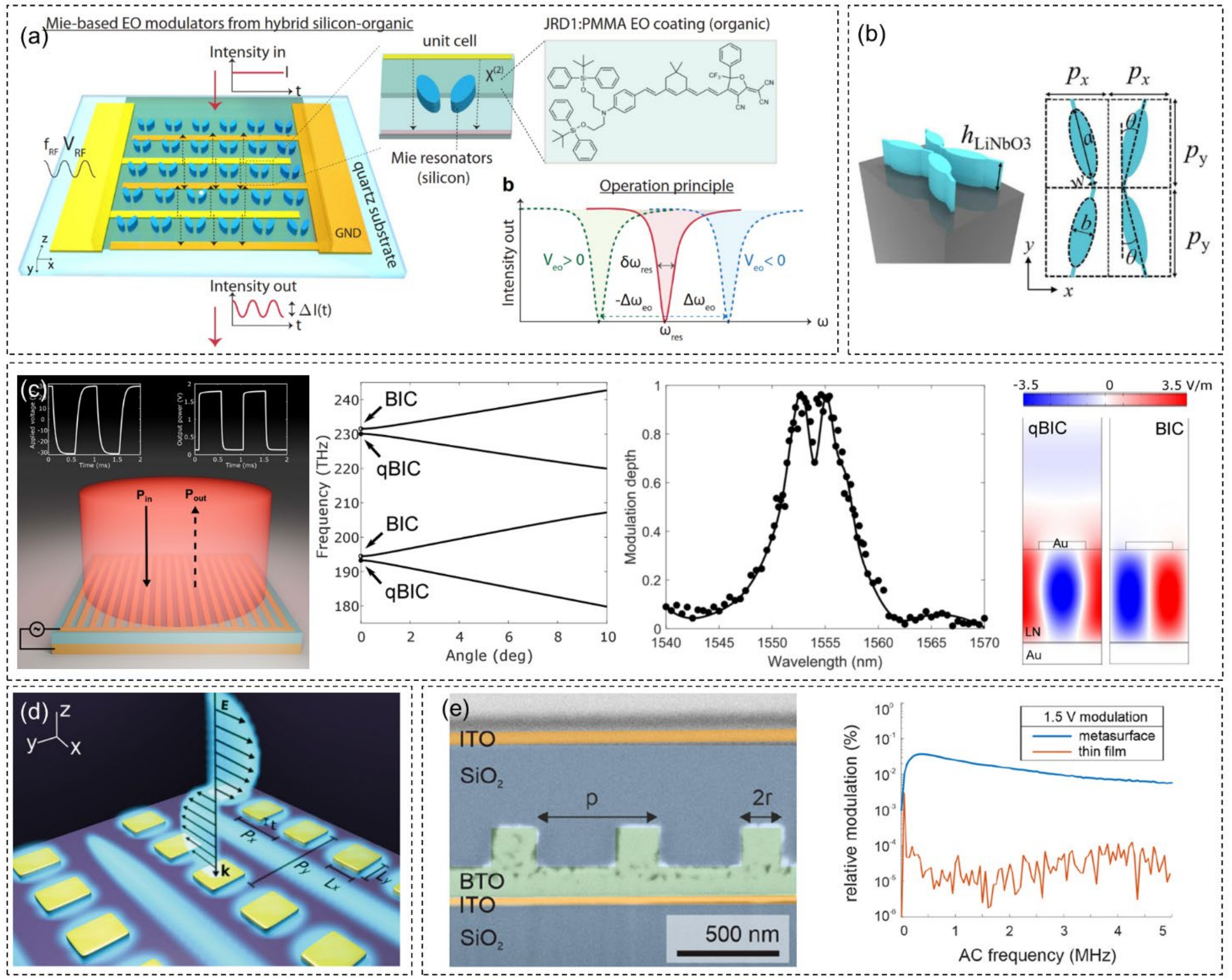

Fig. 8. EO metasurfaces using radiation-suppressed resonances. (a) Free-space electro-optic modulators based on hybrid silicon-organic qBIC. Reproduced with permission[104]. Copyright 2022 Nature Publishing Group. (b) Tunable LN metasurface based on qBIC. Reproduced with permission[165]. Copyright 2024 Science China Press. (c)Tunable qBIC LN metasurface together with the corresponding dispersion curves. Reproduced with permission[74]. Copyright 2025 American Chemical Society. (d) Plasmonic metasurface with a high $Q$ of 2340 in the telecommunication C band by exploiting SLRs. Reproduced with permission[172]. Copyright 2021 Nature Publishing Group. (e) BTO metasurface using SLRs, together with the frequency dependence of optical modulation amplitude on driving AC frequency. Reproduced with permission[173]. Copyright 2024 American Chemical Society.

**SLRs**

SLRs arise in periodic nanoparticle arrays from the hybridization between LSPR and diffractive orders near RA[163]. The RA condition for a 2D lattice under incidence angle $\theta$ can be expressed as:

$$\lambda_{RA} = \frac{n_{eff}\Lambda}{\sqrt{i^2 + j^2}}(1 \pm sin\theta) \tag{19}$$

where $\Lambda$ is the lattice period, $n_{eff}$ is the refractive index of the surrounding medium, and $(i, j)$

denotes the diffraction orders. When $\lambda_{RA}$ spectrally overlaps with the single-particle LSPR, radiative coupling mediated by the diffracted waves forms a collective resonance with reduced radiative damping and a narrowed linewidth. The resulting SLR line shape is often well described by a Fano-type interference model:

$$F(\varepsilon) = \frac{(\varepsilon + q)^2}{\varepsilon^2 + 1}, \qquad \varepsilon = \frac{\omega - \omega_0}{\Gamma/2} \tag{20}$$

where $q$ is the Fano asymmetry parameter and $\Gamma$ is the resonance linewidth[174,175].This collective coupling reduces radiative damping and yields narrow spectral features with steep dispersion, which is attractive for enhancing the efficiency of active modulation.

Bin-Alam et al.[172] demonstrate a plasmonic metasurface with a record $Q$-factor of 2340 at telecom wavelengths (~1550 nm) by exploiting SLRs, exceeding the previous plasmonic record by an order of magnitude (**Fig. 8d**). Through systematic investigation, they identify three key factors limiting observed $Q$-factors: nanostructure polarizability (balancing LSPR-SLR spectral detuning against Ohmic losses), array size (larger arrays support higher-$Q$ standing wave modes), and spatial coherence of the probe beam (coherent illumination doubles the measured $Q$ versus incoherent sources). These design principles are broadly applicable across lattice geometries and particle shapes, reopening plasmonics as a viable platform for high-$Q$ nanophotonic applications such as sensing, nanolasing, and nonlinear optics.

Despite this advantage, SLR-based EO metasurfaces exploiting intrinsic Pockels materials remain relatively scarce. A representative demonstration is provided by Weigand et al.[173], who implemented a nanoimprinted sol−gel-derived polycrystalline BTO metasurface (**Fig. 8e**) and utilized an SLR near 633 nm to enhance Pockels-effect modulation. By biasing the device through a transparent ITO sandwich electrode and operating on the steep resonance slope, they achieved low-voltage transmission modulation with modulation frequencies up to ~5 MHz. Owing to the resonant enhancement ($Q$ = 200), the modulation amplitude was increased by ~600× compared with an unpatterned BTO film, suggesting a scalable route toward large-area free-space EO modulators.

In comparison, SLR tunability has been more frequently explored using electrically driven refractive-index switching in liquid crystals. Sharma et al.[176] integrated a gold nanoantenna array into an liquid crystals cell and strengthened the SLR dip by electrically reorienting the liquid crystals environment, achieving >50% modulation at voltages down to ~2 V, while the LSPR response remained less affected. A.P.van Heijst et al.[177] further demonstrated reversible tuning of SLR dispersion in an aluminum nanorod array combined with an liquid crystals cell, yielding a resonance shift of $\Delta\lambda \approx 7.44$ nm and a modulation depth of ~50%–90%, and additionally enabling voltage-controlled switching of an ITO quasi-guided mode under TE polarization.

Overall, these results indicate that while SLRs have not yet been widely adopted in Pockels-based EO metasurfaces, their high-$Q$ spectral response provides a practical mechanism for low-voltage modulation, and liquid crystals-based implementations illustrate their broader applicability for reconfigurable nanophotonic devices.

### 3.5 Hybrid-mode EO Metasurface

Hybrid-mode EO metasurfaces refer to EO metasurfaces in which two or more resonant mechanisms are intentionally co-designed and coupled—such as plasmonic resonances, GMRs, BIC/qBIC states, and FP cavity—to enhance electrically induced refractive-index modulation. Compared with single-mode metasurfaces (*e.g.*, purely Mie-, SPPs-, or GMR-based designs), hybrid-mode schemes can simultaneously leverage strong near-field localization and spectrally sharp responses, thereby improving modulation efficiency by operating on the steep slope of narrow spectral features. In addition, hybridization often provides extra degrees of freedom for multifunctionality (*e.g.*, polarization selectivity, dual-band operation, and form-factor integration). These benefits, however, typically come with increased structural complexity and stronger sensitivity to fabrication tolerances, as well as potential penalties from metallic absorption and thermal drift.

Within this framework, Zhang et al.[178] demonstrated a free-space EO metasurface modulator based on a hybrid "perfect absorber + bimodal resonance anti-crossing + critical coupling + EO polymer" concept(**Fig. 9a**). Using a MIM Au grating with an embedded EO polymer, two TM MIM modes undergo anti-crossing interference that suppresses radiation leakage, increases the effective $Q$ factor, and enables critical coupling, leading to near-perfect absorption (−27 dB, 99.8%) with $Q = 113$ at 1650 nm. Electrical tuning of the polymer refractive index shifts the resonance and yields intensity modulation on the steep spectral slope, achieving 9.5 dB modulation depth at ±30 V and a measured 3-dB bandwidth of 1.25 GHz, with the speed mainly limited by contact resistance. In a related direction emphasizing system integration, Zhang et al.[179]( **Fig. 9b**) reported a plug-and-play metafiber modulator integrated on a single-mode fiber end face, where the EO response is enhanced through a hybrid "plasmonic nanoeye resonance + guided-mode resonance (diffraction-enabled) + FP cavity interference + EO polymer" scheme. This multi-resonance coupling supports dual-band operation in the telecom O band (~1283 nm) and S band (~1500 nm) and enables amplitude modulation up to ~1 GHz at ±9 V, highlighting the utility of hybrid-mode designs for compact fiber-integrated modulators.

Hybridization has also been exploited to increase functionality and programmability. Ju et al. [180] demonstrated a LN metasurface spatial light modulator enabled by a hybrid "plasmonic qBIC + GMR + Pockels effect" mechanism on an LNOI platform(**Fig. 9c**). TM excitation accesses a plasmonic qBIC, whereas TE excitation accesses a GMR within the same geometry, providing polarization-selective resonance channels across 1480–1620 nm. By tuning the LN refractive index via the Pockels effect and operating on the steep resonance slopes, the device achieves a resonance depth of ~51%, an insertion loss of −2.8 dB at 1510 nm, and dynamic intensity modulation up to 190 MHz with a peak SNR of ~49 dB at 65 MHz, indicating the potential of qBIC/GMR hybridization for multifunctional, polarization-dependent spatial modulation.

The hybrid modes will bring more design freedoms for the EO metasurface. The high-*Q* resonances can be tuned by coupling a qBIC metasurface with a FP cavity through Fano interference—an underexplored mechanism that decouples *Q*-factor enhancement from fabrication precision of structural asymmetry[181] (**Fig. 9d**). Using coupled-mode theory, the authors analytically reveal that *Q* factor, Fano asymmetry, and phase singularity are all controllable via cavity phase (gap thickness, scaling ratio, asymmetry, refractive index), predicting periodic transitions between Lorentzian and Fano line shapes. Simulations confirm *Q* up to $2.9\times10^4$ near the phase singularity. Near-infrared experiments verify Fano inversion across the singularity and achieve 732 $\mathrm{nm\cdot RIU^{-1}}$ refractive-index sensitivity—among the highest across all metasurface, photonic crystal, and plasmonic platforms. This phase-engineered qBIC–cavity interference paradigm offers a compact, analytically tractable building block for strong light–matter interaction beyond sensing, extendable to multilayer and moiré photonic architectures.

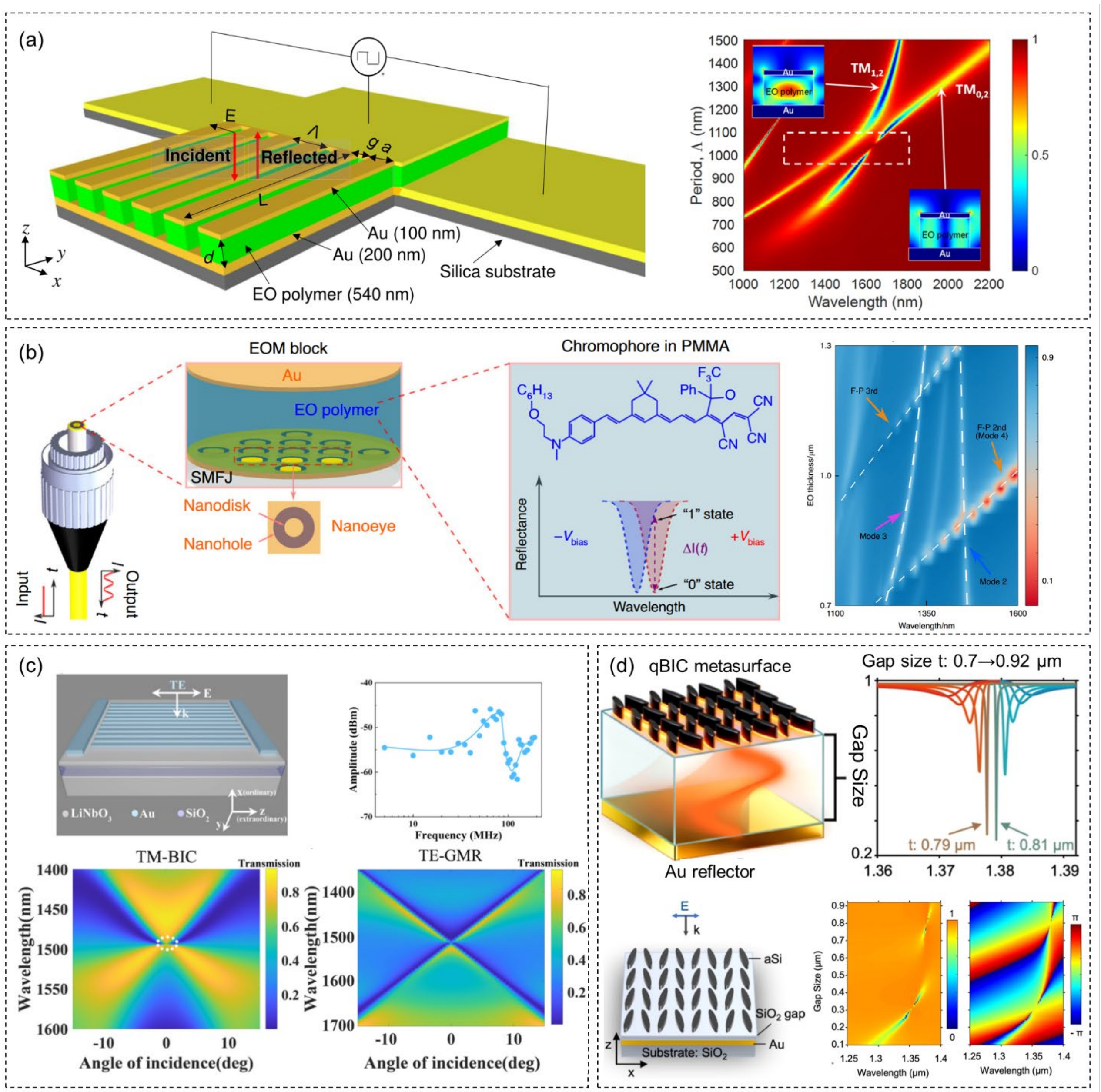

Fig. 9. EO metasurfaces using hybrid-mode resonances. (a) Free-space reflective metasurface modulator based on an MIM Au grating and EO polymer, showing the measured resonant reflectance spectrum ($\Lambda$ = 1080 nm) with near-perfect absorption. Reproduced with permission[179], under a Creative Commons Attribution 4.0 International License.(b) Plasmonic metafiber EOM integrated on a single-mode fiber end face with a three-layer stack (nano eye metasurface/EO polymer/Au electrode), together with the fabrication flow (FIB patterning, polymer spin-coating, and Au deposition). Reproduced with permission[178]. Copyright 2023 Nature Publishing Group. (c) TM-polarized resonance mechanism of an LNOI metasurface modulator, including the device schematic and simulated mode/field profiles of a plasmonic qBIC. Reproduced with permission[180]. Copyright 2023 The Royal Society of Chemistry. (d) Hybrid dielectric-metallic system that couples a qBIC with a FP cavity mode, achieving tunable high Q resonances through Fano interference. Reproduced with permission[181]. Copyright 2026 American Chemical Society.

Overall, these studies suggest three representative hybrid-mode strategies with different strengths:

(i) perfect-absorber/critical-coupling designs combined with multimode interference and EO materials, which are advantageous for high-contrast and potentially high-speed intensity modulation; (ii) plasmonics coupled with GMR/BIC resonances and Pockels-active materials, which are well suited for narrowband, high-sensitivity modulation and programmable/polarization-multiplexed functions; and (iii) plasmonics combined with quasi-BIC engineering and electro-thermal tuning, which enables low-voltage and large spectral tunability at the cost of modulation speed. Future progress will likely rely on co-optimizing resonance line shapes, optical loss, and electrical RC constraints, together with improved materials and electrode designs, to simultaneously achieve high efficiency, high speed, and scalable integration in hybrid EO metasurfaces.

# 4. Performance Comparison and Application of Different EO Metasurface

## 4.1 Performance Comparison of EO Metasurfaces Based on Different Mechanisms

Among the six representative EO metasurface mechanisms listed in **Table 3**—namely Mie, qBIC, SLR, GMR, FP, and plasmonic resonances—devices exhibit pronounced differences in their capabilities for phase, amplitude, spectral, and polarization modulation, as well as in key performance metrics such as modulation depth, response speed, bandwidth, optical loss, and power consumption. These differences fundamentally originate from the distinct local field confinement schemes, effective mode volumes, and resonance $Q$ factors associated with each structural mechanism.

Dielectric metasurfaces based on Mie resonances feature low optical loss and relatively broad bandwidth, and can readily support continuous 0–2π phase modulation. However, their EO modulation depth is usually limited by the intrinsic refractive-index tunability of the constituent materials[182–184].

qBIC/Fano-resonant structures dramatically amplify small refractive-index perturbations through ultrahigh-$Q$ resonances, enabling extremely high amplitude and phase modulation contrast at low driving voltages. Nevertheless, their operational bandwidth is intrinsically ultranarrow, and their performance is highly sensitive to fabrication imperfections and environmental perturbations[185–187].

SLR–based structures represent typical nonlocal high-$Q$ resonant modes that simultaneously possess relatively large mode volumes and strong field enhancement. They exhibit exceptional sensitivity to refractive-index variations and are therefore well suited for high-sensitivity narrowband modulation. However, their large-area periodic arrays tend to introduce substantial parasitic capacitance, which may limit modulation speed[188,189].

GMR metasurfaces extend the effective optical path length through waveguide-mode coupling, enabling large phase modulation with comparatively low propagation loss. These characteristics make them particularly suitable for low-loss, large-aperture modulation and beam-shaping applications[190–192].

FP cavity–enhanced metasurfaces significantly magnify EO effects via multiple round trips of light inside the cavity, allowing high-contrast amplitude modulation at low driving voltages. However, their spectral response is generally narrowband due to the intrinsic nature of cavity resonances[193–195].

Plasmonic and hybrid-integrated structures, relying on extremely small mode volumes and ultrastrong local field enhancement, enable ultrahigh-speed modulation in the GHz regime and beyond. This advantage, however, is accompanied by substantial insertion loss and elevated power consumption caused by unavoidable Ohmic losses in metals [196–198].

Overall, it is difficult for any single modulation mechanism to simultaneously optimize high modulation depth, fast operating speed, wide bandwidth, low optical loss, and low power consumption. The ultimate device performance is determined by a comprehensive trade-off among resonance characteristics, material EO coefficients, and electrode-driving efficiency, as summarized in **Table 3**.

**Table 3**. Performance Comparison of Different EO Metasurface Mechanisms

| Material platform | Resonance mode | Speed bandwidth | Modulation efficiency/depth | Operational voltage | Q factor | Linewidth (nm) | Ref |
|---|---|---|---|---|---|---|---|
| LNOI | qBIC/GMR | 1GHz | $0.015V^{-1}$ | 10 V | 8000 | <0.2 | [85] |
| LN | Plasmonic resonance | 0.8MHz | $0.016V^{-1}$ | ±25 V | 70 | ~22 | [83] |
| LN | GMR | 13.5MHz | $0.042V^{-1}$ | ±10 V | 30 | ~30 | [39] |
| LN | Mie resonance | 2.5MHz | <1 Vpp | 10 Vpp | 129 | ~6 | [134] |
| LN | GMR | 1.6GHz | 0.001 nm/V | ±200V | 1400 | 1.1 | [84] |
| OEO | Mie/qBIC | 3GHz | $0.0067V^{-1}$ | 60 V | 550 | ~2.9 | [104] |
| OEO | GMR | 50MHz | $0.0046V^{-1}$ | ±80 V | – | – | [32] |
| OEO | GMR | 0.38GHz | $0.25 V^{-1}$ | ±1V | 665 | 2.3 | [101] |
| OEO | Plasmonic resonance | 1.25GHz | $0.030V^{-1}$ | ±30 V | 113 | ~14.6 | [179] |
| BTO | hybrid Mie/SLR | 5MHz | $0.014V^{-1}$ | 5 V | 200 | ~3.2 | [173] |
| BTO | Plasmonic resonance | 20MHz | $3.75\times10^{-4}V^{-1}$ | 4 V | – | – | [199] |
| PZT | SPPs | 200kHz | $0.027 V^{-1}$ | 15V | – | – | [96] |

Modulation efficiency: relative modulation per applied Volt.

## 4.2 Applications of EO Metasurfaces

Given the performance advantages delineated above, EO metasurfaces demonstrate distinct superiority across a broad spectrum of emerging photonic systems. Their application domains can be systematically categorized into six principal areas: beam steering and wavefront control; high-speed optical modulation; tunable spectral devices; holography and dynamic displays; quantum photonic applications; and optical computing and neuromorphic photonics.

### Beam Steering and Wavefront Control

EO metasurfaces engineered for beam steering and spatial light control assume a pivotal role in a diverse array of advanced photonic devices, including dynamic metalenses[147], OPAs[142,157,200,201], and spatial light modulator[39,84]. These applications impose stringent performance requirements, specifically demanding large optical apertures, high spatial resolution, low propagation losses, broadband operational capability, and continuous phase programmability spanning the full 0–2π range.

All-dielectric and hybrid metasurface architectures have demonstrated exceptional suitability for meeting these demanding specifications by enabling electrically tunable phase gradients that facilitate dynamic wavefront shaping in real time. Among the various architectural paradigms, metasurfaces founded on Mie resonances[104], FP resonance, and GMR[201] have emerged as particularly optimal platforms for beam steering and wavefront control applications.

Representative demonstrations include electrically tunable dielectric metalenses and EO beam-steering metasurfaces, enabled through the integration of active semiconductor quantum wells or tunable dielectric resonators[202,203].

### Amplitude Modulation

High-speed EO modulation on amplitude[39,82,204]—fundamentally critical for optical communication modules, integrated photonic interconnects, and optical coherence manipulation[205]—demands ultrafast response characteristics, superior modulation efficiency, and compact device architectures with minimized RC time constants. For this application class, the most optimal structural configurations comprise plasmonic metasurfaces, plasmonic-dielectric hybrid architectures, and compact GMR-based modulators.

Plasmonic and hybrid metasurface platforms exploit intense local field enhancement mechanisms and deep subwavelength modal confinement, resulting in exceptionally low parasitic capacitance and enabling modulation bandwidths spanning MHz–GHz regimes and extending to tens of GHz. Their ultra-small active volumes facilitate extremely high-density integration capabilities for fast pixelated modulators and high-speed OPAs. GMR-based EO modulators demonstrate strong refractive-index sensitivity and extended photon lifetimes, enabling substantial phase shifts or amplitude modulation at reduced drive voltages—particularly when integrated with EO materials such as LN, BTO, or OEO polymers.

Recent experimental realizations have successfully demonstrated sub-nanosecond EO modulation, deep amplitude modulation exceeding 90%, and qBIC-enhanced ultrafast EO switching. These achievements collectively underscore the exceptional suitability of these metasurface platforms for next-generation optical communication infrastructure and advanced beam-steering systems[206].

### Tunable Spectral Devices

Tunable spectral devices—including wavelength-selective filters, spectral modulators, and refractive-index sensors—necessitate high $Q$-factors, robust spectral selectivity, substantial modulation contrast, and stable modal confinement characteristics. The most appropriate metasurface architectures for these applications encompass GMR metasurfaces, qBIC metasurfaces, SLR metasurfaces, and FP cavity-enhanced designs.

GMR and qBIC metasurfaces support exceptionally high $Q$-factors ranging from $10^3$ to $10^5$, enabling ultra-narrowband filtering capabilities and substantial electro-optically induced wavelength shifts[85,160,207]. SLR metasurfaces stabilize narrow spectral linewidths through lattice-coupled resonance mechanisms, delivering outstanding performance in high-resolution sensing applications[172]. FP-enhanced metasurfaces leverage electromagnetic wave longitudinal coupling to minimize required tuning voltages while simultaneously achieving high spatial and spectral resolution[208,209].

These platforms have demonstrated continuous tunability spanning tens of nanometers, pronounced refractive-index sensitivity, and low-voltage spectral switching functionality. These capabilities collectively illustrate their technical superiority for spectrally selective EO devices[210].

## Quantum Photonics

Quantum photonics imposes exceptionally stringent requirements on optical propagation losses, phase manipulation precision, and compatibility with single-photon quantum states. Consequently, all-dielectric metasurfaces based on Mie-resonant, GMR, and qBIC represent the most suitable platform architectures for these applications.

Dielectric metasurfaces provide extremely low absorption characteristics and high-fidelity phase control, enabling deterministic wavefront shaping for single-photon states, quantum mode sorting operations, and quantum interference engineering. qBIC metasurfaces deliver extremely high field enhancement factors, substantially boosting nonlinear processes such as spontaneous parametric down-conversion and spontaneous four-wave mixing, thereby serving as compact quantum light sources with dynamically tunable properties[85,211]. EO metasurfaces integrated with LN or BTO further facilitate reconfigurable quantum gates and tunable single-photon phase shifters.

State-of-the-art review articles comprehensively summarize metasurfaces as critical enabling technologies for quantum state engineering, entangled-photon generation, and compact quantum sensing systems[69,212].

## Optical Computing and Neuromorphic Photonics

Optical computing architectures and neuromorphic photonic systems necessitate metasurfaces capable of reconfigurable phase and amplitude modulation, analog weight tuning, high-speed response characteristics, and massively parallel computational processing[27,213–216]. The most suitable structural platforms for these applications comprise GMR metasurfaces, FP cavity-enhanced EO metasurfaces, and plasmonic metasurfaces.

GMR and FP cavity-assisted metasurface architectures provide extensive, continuous phase tunability with minimal power consumption, making them ideal for implementing tunable synaptic weights in optical neural network implementations. Plasmonic metasurfaces, despite exhibiting higher optical losses, offer extreme device miniaturization, high speed, and nonlinear activation-like behavior, thereby enabling dense computational architectures. All-dielectric phase metasurfaces support high-efficiency linear transformations for large-scale optical matrix multiplication operations.

Recent literature highlights EO metasurfaces as powerful technological platforms for scalable photonic neural networks and hybrid photonic-electronic computing accelerators[69].

# 5. Challenges and Future Perspectives

This review has systematically presented EO metasurfaces encompassing diverse modulation mechanisms, including Mie resonances, qBICs, SLRs, GMR, FP cavities, and plasmonic hybrid structures. Synthesized with the comparative analysis of device performance metrics and application suitability presented in previous sections (**Table 3**), it becomes evident that these resonance mechanisms exhibit distinct advantages across multiple performance dimensions: phase, amplitude, spectral, and polarization control capabilities[75], as well as modulation depth, operational speed, spectrum bandwidth, optical losses, and power consumption characteristics.

However, the principal bottleneck constraining translation of these demonstrated physical advantages into practical, engineerable devices and systems stems from limitations imposed by fabrication technologies and integration strategies. This technological gap represents the critical barrier preventing widespread adoption and commercialization of EO metasurfaces despite their promising theoretical performance.

## 5.1 Challenges of EO Material Requirements and EO Metasurface Fabrication

From a fabrication perspective, nanolithography precision, dry etching quality, thin-film transfer reliability, and electrode fabrication constitute the technological foundation for realizing high-performance EO metasurfaces[217]. Taking LNOI platforms as a representative example, high-$Q$ qBIC or higher-order Mie-resonant EO metasurfaces typically require steep vertical sidewalls (aspect ratios of 5:1–10:1 for sub-100-nm lateral features) and ultralow surface roughness (RMS < 1 nm), placing stringent demands on EBL resolution, mask quality, etching chemistry, and ICP-RIE process control [94]. High-quality LN nanogratings fabricated through combined FIB milling and ICP-RIE can sustain sharp qBIC resonances with $Q$ factors exceeding $10^4$; however, FIB suffers from limited throughput and $Ga^+$ implantation damage that degrades surface quality for shallow features[218]. These results underscore that fabrication precision directly determines achievable $Q$ factors and EO modulation sensitivity.

This sensitivity becomes particularly acute for high-$Q$ BIC/qBIC structures, where the pursuit of larger $Q$ is intrinsically at odds with fabrication tolerance and long-term stability. While high-$Q$ metasurfaces dramatically amplify EO effects through field enhancement, they simultaneously amplify sensitivity to geometric imperfections and environmental fluctuations. For low-$Q$ BIC designs ($Q \sim 200$), a geometric standard deviation of ~0.9 nm reduces $Q$ by 40–110 from the as-designed value [219,220]; for high-$Q$ qBIC structures ($Q > 10^4$), the same order of geometric deviation can cause $Q$ to collapse by orders of magnitude, as the inverse-square scaling of radiative $Q$ with the asymmetry parameter leaves negligible margin for fabrication error. This tension necessitates systematic co-optimization of structural design parameters—including fabrication-robust unit cell geometries and shallow-etch strategies that minimize field overlap with imperfect sidewalls[219,220]—material selection criteria, and process windows. Alternative approaches, such as etchless LN metasurfaces that avoid direct LN patterning, offer a promising route to circumvent these requirements while still achieving moderately high $Q$ factors[221].

Beyond nanoscale patterning, the preparation and poling of high-$Q$ EO thin films underpin the performance of BTO-, PZT-, and OEO-based metasurfaces. The deposition method and processing conditions govern the growth orientation, residual stress, surface roughness, crystallinity, thickness uniformity, and defect density of the resulting films, which in turn dictate their electro-optic coefficients.

**BTO**. As-grown BTO films contain four orthogonal domain states (c-axis variants at 0°, 90°, 180°, and 270°); the net Pockels coefficient is zero in the unpoled state and depends on the fractional volume of each domain orientation[76,222]. Ferroelectric domain pinning leads to polarization drift under prolonged operation, and achieving complete single-domain poling in ultrathin BTO films (<100 nm) remains extremely difficult, with residual polarization inhomogeneity inducing unwanted birefringence. Partially poled devices are susceptible to performance degradation from bias perturbations, though fully poled devices are resilient[90]. Secondly, heterogeneous integration of epitaxial BTO onto silicon wafers demands precise stress management to prevent cracking and maintain crystalline quality[90]. Soft nanoimprint lithography (SNIL) offers an etch-free route to wafer-scale BTO photonic device fabrication[88]. Thureja et al.[223] demonstrated that single-crystalline BTO thin films exfoliated via spalling preserve bulk EO properties, yielding $r_{33}$ = 160 pm/V in single-domain regions with projections up to $r_{42}$ = 1980 pm/V under unclamped conditions—thereby exceeding commercially available TFLN. Moreover, BTO also faces the constraint of a low Curie temperature (Tc ≈ 120°C for bulk; up to ~220°C under epitaxial strain[224]), which limits thermal processing windows, and generates non-volatile etching byproducts ($BaF_2$, $BaCl_2$) that resist standard post-etch cleaning[225].

**PZT**. Chemical solution deposition (CSD)-grown PZT polycrystalline films contain abundant in-plane ferroelectric domains, leading to device-to-device modulation depth variations as large as a factor of two. Under strong applied electric fields, domain reorientation—rather than the intrinsic Pockels effect—dominates the modulation response, imposing an unusual low-frequency cutoff at ~200 kHz, far below the intrinsic THz-level bandwidth of the Pockels effect[96]. To obtain a significant electro-optic response, a poling step is usually performed, i.e., applying 60–80 V (≈150 kV $cm^{-1}$) for 1 h at room temperature, followed by several hours of stabilization time[95]. Furthermore, the inherent lead content raises concerns regarding CMOS foundry compatibility and environmental regulations, restricting PZT integration to back-end-of-line processes.

**OEO**. OEO polymers provide substantially larger EO coefficients than LN (poled bulk films: $r_{33}$ up to 1100 pm/V[100]; in-device values: 150–390 pm/V depending on platform and electrode geometry) along with femtosecond intrinsic response times, low dielectric constants ($\varepsilon < 7$), and direct spin-coating deposition. However, their practical deployment is severely constrained by thermal and photochemical instability. Chromophore relaxation above ~85°C causes rapid $r_{33}$ decay, incompatible with commercial packaging requirements that mandate stable operation at 85°C for >2000 h. High-intensity optical illumination induces two-photon absorption and photobleaching, undermining long-term reliability[226]. Furthermore, at high chromophore concentrations, intermolecular electrostatic interactions drive aggregation, paradoxically reducing the effective EO response by disrupting the acentric order achieved during poling[100]. Recent cross-linkable chromophore systems have made substantial progress: HLD1/HLD2 achieves Tg ~174° C after cross-linking with >99% $r_{33}$ retention over 500 h at 85°C, while BAH-X systems demonstrate >2000 h thermal stability at 85°C under nitrogen with cross-linked $r_{33}$> 650 pm/V[227]. These results represent significant steps toward telecom reliability standards, though long-term photostability and humidity resistance remain to be fully validated.

In summary, these four platforms form a complementary hierarchy: LN offers the most mature and reliable EO platform—low propagation loss, broadband flat frequency response without dispersion, no poling complications, and established wafer-scale availability—but its modest $r_{33}$ (~30 pm/V) limits per-pixel modulation depth without resonant enhancement, and its reliance on crystal slicing and bonding rather than direct epitaxy constrains integration flexibility; BTO provides the highest inorganic effective EO coefficient with silicon process compatibility, but demands careful poling management and suffers from frequency dispersion above ~100 MHz; PZT offers competitive EO coefficients with excellent thermal and chemical stability at low cost, but is limited by multi-domain effects and lead content; OEO polymers deliver the largest absolute EO response with the lowest dielectric constant—enabling low $V_{\pi}$ and high RC-limited bandwidth—but face fundamental reliability constraints from thermal instability and photodegradation. The convergence of advanced wafer bonding and etchless patterning (for LN), strain-engineered epitaxial growth (for BTO), domain-controlled processing (for PZT), and theory-guided molecular design (for OEO) will determine which platform—or hybrid combination—ultimately enables commercially viable, wafer-scale EO metasurface systems.

### 5.2 Heterogeneous Integration Challenges

Regarding thin-film integration, reliable incorporation of high-EO-coefficient crystalline materials—such as LN and BTO—onto silicon or glass substrates constitutes a prerequisite for successful co-integration of EO metasurfaces with existing photonic integrated circuits (PICs). Standard integration strategies include wafer bonding (direct, plasma-activated, or adhesive bonding), die-to-wafer (chiplet) bonding, and micro-transfer printing. Notably, micro-transfer printing has enabled the heterogeneous integration of LN coupons onto 200-mm silicon photonics wafers with a 3σ placement accuracy of 1 μm, yielding modulators with $V\pi$ = 3.8 V and bandwidths exceeding 70 GHz[228]. Low-loss heterogeneous integration has been experimentally demonstrated on LNOI–Si and LNOI–$Si_3N_4$ platforms for high-speed EO modulators, providing a feasible technical pathway for integrating EO metasurfaces into mature CMOS-compatible photonic platforms. By contrast, OEO polymers are deposited by solution processing and in-situ electric-field poling, circumventing the need for bonding or transfer altogether. For free-space optical applications based on glass substrates, achieving low-defect, stress-controlled large-area thin-film transfer remains a significant technical challenge for realizing large-aperture EO metasurfaces intended for spatial light control. Addressing this challenge requires the development of novel transfer techniques that can maintain film integrity across cm-scale apertures while minimizing residual stress-induced birefringence and crack formation.

### 5.3 Electrical and Optical Co-Design Considerations

With respect to electrode fabrication and driving-circuit integration, different EO metasurface resonance mechanisms impose distinct electro-optical co-design priorities. High-$Q$ qBIC and GMR structures rely on extended nonlocal arrays—typically with periods of 600–1000 nm at telecom wavelengths—to sustain sharp resonances, which inevitably increases the device footprint and total capacitance. For MIM electrode configurations, the capacitance per unit area can reach tens of pF/mm$^2$ (*e.g.*, ~44 pF/mm$^2$ for a 540-nm-gap EO polymer MIM[179]), and even higher for thin high-ε dielectrics such as LN. In dielectric structures using doped-semiconductor electrodes (*e.g.*, Si), the modulation bandwidth is primarily limited by the electrode resistivity rather than capacitance alone, creating a fundamental trade-off between electrical conductivity and free-carrier absorption. InP-membrane electrodes have recently been shown to break this trade-off, enabling a record 17.5-GHz bandwidth with $Q \approx 100$ and only 0.56 dB insertion loss[229].

Plasmonic-organic hybrid structures employ metallic gratings as both optical resonators and low-resistance electrodes, yielding inherently low RC time constants. However, their plasmonic loss results in significant insertion loss (typically 10–27 dB for free-space metasurface modulators) and limited Q factors (<120). The highest-speed plasmonic metasurface modulator to date achieved 9.5 dB modulation depth at 1.25 GHz bandwidth using a critically coupled bimodal MIM resonance[179], though the measured bandwidth was limited by extrinsic contact resistance and driver impedance rather than the device capacitance, which theoretically supports operation beyond 150 GHz. These results indicate that the design space for EO metasurface modulators is not simply partitioned into "high-$Q$–low-speed" and "low-$Q$–high-speed" regimes; rather, the optimal operating point depends on a multidimensional trade-off among electrode material properties (resistivity versus optical absorption), resonance $Q$ and the associated photon-lifetime bandwidth ceiling, device footprint, and the specific modulation depth, bandwidth, power, and insertion-loss requirements of the target application.

### 5.5 Future Outlook

EO metasurfaces stand at a pivotal juncture: the physical principles are established, early-stage devices have been demonstrated, yet the simultaneous attainment of high modulation depth, wide bandwidth, and low driving voltage—within a single free-space platform—remains elusive. Realizing this convergence will require coordinated advances across three interdependent axes: EO materials (the focus of this section), resonance–electrode co-design, and scalable heterogeneous integration.

**EO materials for metasurfaces.** The ideal EO material for a metasurface pixel must deliver a large effective Pockels coefficient at sub-micron interaction lengths, low absorption at the operating wavelength, and compatibility with nanofabrication processes. Ferroelectric thin films currently lead in maturity: LN offers $r_{33} \approx 30$ pm/V with exceptional optical quality and established wafer-scale availability[230], while monolithic BTO-on-insulator microresonators have demonstrated intrinsic $Q > 10^6$ with 0.32 dB/cm propagation loss[224]. Spalled single-crystalline BTO films yield $r_{33} = 160$ pm/V in single-domain regions, with projections up to $r_{42} = 1980$ pm/V under unclamped conditions[223]—though this $r_{42}$ value represents an upper bound, as clamped conditions relevant to GHz-frequency device operation yield substantially lower effective coefficients. For BTO-on-Si platforms, effective Pockels coefficients up to ~900 pm/V have been achieved under optimal poling, compared with the intrinsic bulk $r_{33} \approx 105$ pm/V. OEO polymers complement these inorganic platforms with $r_{33}$ values reaching 1100 ± 100 pm/V[100]; cross-linkable variants have achieved $r_{33} > 650$ pm/V with Tg ~ 150°C (BAH-X[227]) and $r_{33}$~ 290 pm/V with Tg ~ 174°C (HLD[227]), offering lower-temperature processing and conformal coating on pre-patterned nanostructures.

The choice between crystalline and polymer platforms entails a fundamental trade-off: inorganic films provide long-term stability and high damage thresholds but require high-temperature deposition and challenging etch or transfer steps at sub-10-nm precision, whereas polymers enable facile integration but face photodegradation under sustained illumination[226] and drift under bias. For metasurface pixels specifically, this trade-off is sharpened: the sub-wavelength pixel size demands EO coefficients that are effective over interaction lengths of a few hundred nanometers, where inorganic resonant enhancement via qBIC or Mie modes becomes essential to compensate for modest material $r_{33}$, whereas OEO polymers can achieve sufficient modulation depth without resonant structures but at the cost of long-term reliability.

For BTO, the immediate priorities are poling stability and frequency dispersion management. Ferroelectric domain pinning causes polarization drift under prolonged operation; optimized in-situ poling and annealing protocols are needed to suppress pinning and achieve stable GHz-bandwidth operation. Strain engineering to accommodate lattice mismatch and enable crack-free epitaxial growth on silicon substrates remains essential for wafer-scale integration. The discovery by Chelladurai et al. that certain device geometries can maintain a flat EO frequency response despite material dispersion provides a valuable design principle for next-generation BTO modulators[76]. For metasurface applications, BTO's large effective EO coefficient enables compact pixel designs, but the dispersion of $r_{42}$ above ~100 MHz must be accounted for in broadband beam-steering or switching architectures.

For PZT, the transition from CSD to epitaxial growth techniques that yield single-domain, preferentially oriented films would eliminate parasitic domain reorientation and recover the intrinsic high-speed Pockels response. The development of lead-free PZT dopant variants that preserve the high $r_{33}$ coefficient while meeting RoHS compliance would remove a major barrier to foundry adoption. Plasmonic or resonant mode engineering—such as SPP waveguides—can compensate for the moderate refractive index by concentrating the optical field within the PZT layer, improving the field overlap integral and modulation efficiency[96]. In the metasurface context, PZT's low cost and chemical robustness make it attractive for large-area, multi-pixel arrays where per-pixel speed requirements are modest (*e.g.*, sub-MHz reconfigurable holograms).

For OEO, theory-guided molecular engineering of chromophores with higher glass-transition temperatures through cross-linkable architectures that lock in the poled acentric order is the central strategy. Advanced encapsulation designs that inhibit oxygen diffusion and reduce two-photon absorption cross-sections would improve photostability. The development of high-refractive-index ($n > 2$) chromophores is particularly attractive, as the modulator figure of merit $V_{\pi}L$ scales inversely with $n^3r_{33}$; even moderate increases in $n$ would yield substantial device-level improvements. The charge-blocking-layer strategy demonstrated by Xu et al. represents a critical enabling technique that allows high-concentration chromophores to achieve record poling efficiencies without leakage current degradation[100]. For metasurface pixels, OEO's conformal coating capability is uniquely advantageous for depositing on three-dimensional nanostructure topographies, though the thermal and photostability constraints limit deployment to controlled environments or low-duty-cycle applications.

**Scalable heterogeneous integration.** Translating EO metasurfaces from millimeter-scale laboratory demonstrations to wafer-scale functional devices requires solving heterogeneous integration challenges that are more demanding than those for passive metasurfaces or waveguide modulators individually. The EO thin film must be combined with a high-$Q$ resonant nanostructure and a metal electrode pattern, all while preserving sub-5-nm fabrication precision[231] and maintaining optical quality across the entire aperture. Key milestones have been achieved in adjacent domains: 200-mm CMOS-compatible TFLN modulators with bandwidth > 110 GHz and $V_\pi$ < 3 V at ~50% yield[230], and micro-transfer printing of 217 LN coupons onto silicon photonic wafers with $V_\pi$ = 3.8 V and bandwidth > 70 GHz[228]. However, integrating an EO thin film with a pre-patterned metasurface nanostructure—rather than a planar waveguide—remains at an early stage, and monolithic co-fabrication of EO metasurfaces with electronic driver circuits has not yet been demonstrated. Critical challenges include defect and stress management across heterogeneous interfaces, development of alignment-tolerant transfer bonding processes, and establishment of design rules that balance optical performance with fabrication yield. Wafer-scale DUV patterning of qBIC metasurfaces on 4-inch substrates[232] and 8–12-inch NIL of passive metasurfaces[233,234] provide fabrication infrastructures that could be adapted for EO variants once the integration process matures.

**Resonance–electrode co-design.** The central performance bottleneck of EO metasurfaces is the trade-off among resonance $Q$ factor, modulation depth, and switching speed. High-$Q$ resonances enhance the local field and thus the effective phase shift per volt, but the narrow spectral linewidth inherently limits the modulation bandwidth. Passive all-dielectric metasurfaces have reached $Q > 10^5$ via qBIC and low-contrast BIC[235], yet the highest $Q$ experimentally achieved in an EO-active metasurface remains ~8000[85], because electrode deposition and material interfaces introduce additional loss channels. Closing this gap—toward $Q > 10^4$ in actively tunable devices—would enable sub-volt phase shifts of $\pi$ at individual metasurface pixels, but requires electrode designs that preserve the resonance while providing efficient RF delivery. Unlike waveguide-based modulators where traveling-wave electrodes and coplanar waveguides achieve impedance-matched broadband operation, free-space EO metasurfaces are lumped-element devices whose lateral dimensions are comparable to or smaller than the RF wavelength; the appropriate driving architecture is therefore lumped capacitive driving with resonant RF enhancement and impedance-matching networks at the pixel or sub-array level. A compelling precedent is the InP high-contrast grating modulator[229], which achieves 17.5 GHz bandwidth at $Q$ = 102 with 0.56 dB insertion loss, demonstrating that moderate-$Q$ structures can attain GHz-class speeds with low loss—breaking the conventional dichotomy of high-$Q$/slow versus low-$Q$/fast. Multi-resonance strategies that combine Mie, GMR, SLR, and qBIC mechanisms within a single pixel may offer concurrent improvements in modulation depth and bandwidth, though theoretical proposals for GMR–photonic crystal hybrids with $Q$ ~ 1500 and sub-diffraction mode volumes still await experimental validation.

AI-assisted inverse design is rapidly accelerating the metasurface design phase: deep generative models now produce on-demand structures from optical specifications[236], and physics-informed frameworks achieve sub-second optimization for $10^4$-scale meta-atom arrays[237]. Multi-objective optimization that simultaneously maximizes $Q$, bandwidth, and modulation depth—while incorporating fabrication tolerance constraints[238]—is becoming tractable. However, AI currently covers only the design and simulation stages; compressing the full development cycle will require closing the loop with automated nanofabrication and in-line metrology, which remains an aspirational goal.

## 6. Conclusion

In summary, the next generation of EO metasurfaces will emerge not from any single breakthrough but from the co-optimization of EO materials, resonance–electrode architectures, and heterogeneous integration processes. The field is transitioning from demonstrating that per-pixel EO modulation is physically possible toward engineering devices that simultaneously satisfy the performance requirements of real applications—free-space optical communication, LiDAR beam steering, and computational imaging—where high modulation depth, GHz bandwidth, low driving voltage, and scalable fabrication must all converge within a single platform. The confluence of emerging EO thin-film platforms, sub-10-nm nanofabrication, and AI-driven inverse design is accelerating this transition from laboratory demonstrations toward technology readiness, with commercially viable devices anticipated within the next decade. As EO metasurfaces approach modulation rates in the tens of gigahertz while maintaining sub-wavelength pixel footprints, they offer a pathway to dynamic, spatially resolved control of light–matter interactions at length and time scales inaccessible to conventional bulk modulators—enabling programmable photonic systems, both classical and quantum, for beam steering, computational imaging, and reconfigurable nanophotonic circuits.

**Acknowledgments:** This work was supported in part by the National Natural Science Foundation of China (No. 62405280, 62431025), Leading Innovative and Entrepreneur Team Introduction Program of Zhejiang (2024R01001), Fund of European Union (Grant 101185769–OPTIPATH). All authors have accepted responsibility for the entire content of this submitted manuscript and approved submission.

The authors declare no conflicts of interest regarding this article.

**Data availability.** The data that support the finding of this study are available from the corresponding author upon request.